\documentclass[journal]{IEEEtran}
\pdfoutput=1
\ifCLASSINFOpdf
\else
\fi
\usepackage{tikz}
\usepackage{graphicx}
\usepackage{mathrsfs}
\usepackage{amssymb}
\usepackage{lipsum}
\usepackage[cmex10]{amsmath}
\newtheorem{defn}{Definition}
\newtheorem{theorem}{Theorem}
\newtheorem{lemma}{Lemma}

\newtheorem{proposition}{Proposition}

\begin{document}
%
\title{Variable Packet-Error Coding}

\author{Xiaoqing Fan, Oliver Kosut, and Aaron~B.~Wagner%
\thanks{X. Fan and A.~B.~Wagner are with the School of Electrical
  and Computer Engineering, Cornell University, Ithaca, NY 14853.}
\thanks{O.~Kosut is with the School of Electrical, Computer and
   Energy Engineering, Arizona State University, Tempe, AZ 85287.}
\thanks{{This work was presented in part at the 2013 IEEE International Symposium on Information Theory~\cite{Kosut:13ISIT} and
the 51st Annual Allerton Conference on Communications, Control,
   and Computing~\cite{fan2013polytope}.}}}

\maketitle
\begin{abstract}
\boldmath
We consider a problem in which a source is encoded into $N$
packets, an unknown number of which are subject to adversarial
errors en route to the decoder. We seek code designs for which
the decoder is guaranteed to be able to reproduce the source
subject to a certain distortion constraint when there are no
packets errors, subject to a less stringent distortion constraint
when there is one error, etc. Focusing on the special case
of the erasure distortion measure, we introduce a code design
based on the \emph{polytope codes} of Kosut, Tong, and Tse. The
resulting designs are also applied to a separate problem
in distributed storage.

\end{abstract}


%
\IEEEpeerreviewmaketitle

\section{Introduction}

Consider a communication scenario in which a source sends 
information to a destination over several nonintersecting
paths in a network. These paths could be used to increase
the data rate beyond what would be achievable with a single
path, or they could be used to provide redundancy to allow
the decoder to recover from errors introduced by the network.
It is also possible to simultaneously achieve both goals, 
subject to a tradeoff between the two, which is 
the topic of this paper. In particular, we shall assume 
that some number of paths are subject to adversarial
errors, and we shall seek codes that achieve high data
rates while still ensuring that the encoder can reconstruct
the original message reasonably well in the face of those
errors.

While coding for adversarial errors is a classical 
subject~\cite{yeung2006network}~\cite{cai2006network}, prior work 
in coding theory
seeks to optimize only the worst-case performance
of the code, that is, how well it performs when the
number of errors introduced by the network is the 
maximum. 
For many real systems, however, this approach
is overly pessimistic. Indeed, if the errors are
due to an attack by an adversarial jammer, then the 
system may experience no errors at all in the typical 
case, since the network may only come under attack 
occasionally. We therefore desire
a system that achieves some performance objective
when the maximum number of errors are present while
guaranteeing that a higher level of performance is
achieved when there are fewer, or no, errors.
This is not provided by the conventional approach
to the problem, which is to use maximum distance
separable (MDS) codes with a minimum distance
that exceeds twice the maximum number of possible
errors. For such codes the decoder can fully
recover the source when the maximum number
of errors occurs, but should no errors occur
then the decoder is no better off than if they did.

We seek designs whose performance improves as the
number of errors decreases.
Since prior work has shown that source-channel separation
is not optimal for this problem~\cite{ebad:itw},
it is properly formulated using rate-distortion theory.
We assume that a source sequence in encoded into $N$ packets 
(or messages) at a given rate $R$, at most $T$ of which 
may be adversarially altered by the network. The decoder
receives $N$ packets without knowing which packets
were altered or how many have been altered
(except that it knows that the total number of altered 
packets
does not exceed $T$). The decoder then outputs a reconstruction
of the source. We are given a \emph{distortion measure}
between the source and reproduction, and we seek codes
that guarantee a certain level of distortion when
there are $T$ errors, a lower level of distortion when
there are $T-1$ errors, and so on.

In this paper we shall focus exclusively on the
\emph{erasure} distortion measure: the per-letter
distortion is zero if the source and reconstruction
symbols agree, one if the reconstruction symbol is
a special ``erasure'' symbol, and infinity otherwise.
Thus there is an infinite penalty for guessing a source
symbol incorrectly, and the decoder should output 
the erasure symbol for any source symbol about which
it is unsure. Assuming there are no errors in the 
reconstruction, the distortion of a string is then
the fraction of erasures in the reconstruction. 
The erasure distortion measure is reasonable for a wide array 
of physical sources. For audio and video, it is typically 
possible to interpolate over unknown samples, pixels, or 
frames at the receiver. Similarly, humans can often recover a
natural language source when some of the characters have
been erased~\cite{chou2006rate}. Even executable computer code,
which is typically viewed as being unamenable to
lossy compression, is suitable to compression under
the erasure distortion measure: execution of the program
at the decoder could simply pause whenever it reached
an erasure and wait for further information, without ever
executing incorrect instructions. Focusing on the erasure
distortion measure is also a useful simplifying assumption
when considering 
new problems, akin to the way that the
binary erasure channel is a good starting point in the
study of modern coding theory~\cite{richardson2008modern}.

For this problem we provide a code construction that is
inspired by the \emph{polytope codes} introduced by
Kosut, Tong, and Tse~\cite{KosutTongTse:14IT} 
in the context of network
coding with adversarial nodes. Polytope codes are
similar to linear maximum distance separable (MDS)
codes but with an added feature: for a certain number
of errors, which exceeds the decoding radius of the
code, it is possible to always decode some of the 
codeword symbols even though it is not possible to
decode all of them. This is to be contrasted
with conventional MDS codes, for which in general none of the
coded symbols can be decoded unless they all can.
This ``partial decodability'' property will be
crucial in our use of polytope codes.
Our construction of polytope codes departs significantly
from that of Kosut, Tong, and Tse, and is arguably more
transparent. Nonetheless, we shall still call them polytope
codes to emphasize their connection to this earlier work.

The problem studied here can be viewed as an instance
of a  ``large-alphabet'' channel. In classical studies of 
channel capacity, the channel law is held fixed and the 
blocklength is permitted to grow without bound (e.g.~\cite{cover2012elements}). 
In the case of discrete memoryless channels with finite alphabet,
this model well captures the practical regime in which the
blocklength is much bigger than the number of channel inputs
or outputs. While this model has proven to be very successful,
the asymptotic that it considers is not always the right one.
For the problem 
in which a sender sends data over several independent paths in 
a network, some of which may alter the data adversarially
en route, the ``blocklength'' is naturally viewed as the
number of distinct paths, which is generally small, while 
the ``alphabet'' is the number of distinct messages
that can be sent on one path, which is generally very large.
Thus the appropriate model is in some sense dual to the 
classical one: the blocklength is fixed while the input and 
output alphabet sizes are permitted to grow without bound,
as is done in this paper.
Such channels have arisen in network coding~\cite{ho2008network}, 
although many fundamental Shannon-theoretic questions 
about them are not well understood. One notable exception
is that, as alluded to earlier, 
source-channel separation is known to be optimal for such 
channels if the source is Gaussian and the distortion measure 
is quadratic or if the source is Bernoulli and the distortion 
measure is Hamming distance but not, in general, if the source 
is binary and the distortion measure is erasure 
distortion~\cite{ahmed2012erasure}. 
Thus we already know that such channels behave differently
from conventional ones. We call communication over such
channels \emph{packet-error} (or \emph{path-error}) \emph{coding} (PEC).

In this paper, we are interested in packet-error coding
in which the number of packet errors is variable and
a single code simultaneously provides different performance
guarantees depending on the number of packet errors. We
call this \emph{variable packet-error coding} (VPEC).
VPEC is closely related to the
multiple descriptions (MD) problem~\cite{goyal2001multiple} in network
information theory. The difference is that in the MD
problem each message is either received correctly or
not received at all; the network does not introduce
errors. The MD problem has received considerable 
attention~\cite{gamal1982achievable,goyal2001multiple,puri1999multiple} 
since it was introduced, including 
the special case in which the distortion measure is 
erasure~\cite{ahmed2012erasure}.
Allowing the adversary to introduce
errors instead of erasures seems to significantly alter
the problem, however. In particular, although techniques
from coding theory have been successfully applied to 
the MD problem~\cite{puri1999multiple}, the polytope codes that 
shall prove so effective here do not appear
to be useful for the MD problem.

Having developed the polytope code constructions for the VPEC problem, we subsequently apply essentially the same codes to the  distributed storage system (DSS) problem in the presence of an active adversary. In a DSS, a file is stored across multiple storage nodes in a redundant fashion so as to recover from node failures. Beginning with Dimakis \emph{et al.}~\cite{DimakisEtal:10IT}, there has been considerable recent interest in applying techniques from network coding to the DSS problem. The problem has also been studied when several of the storage nodes are controlled by a malicious adversary~\cite{DikaliotisDimakisHo:10ISIT,
PawarEtal:11IT,RashmiEtal:12ISIT,
OggierDatta:11P2P,
SilbersteinRawatVishwanath:12Allerton}.

Unlike the network coding problem originally studied for polytope codes~\cite{KosutTongTse:14IT}, in which the network topologies can be arbitrary, the DSS problem yields highly constrained network topologies that are in fact similar to the one-hop network of the VPEC problem. That is, one is confronted with many data packets, some of which may be adversarially corrupted, and trustworthy packets must be identified. This similarity allows the use of the same polytope code constructions, and the partial decodability property will again be critical.

The rest of the paper is organized as follows. Section \ref{problem} describes the VPEC problem in detail and states the main theorem. Polytope codes are then defined in Section \ref{sec:Polytope Codes} and used to prove the main theorem in Section \ref{sec:proof of theorem}. We prove a partial optimality result for polytope codes in Section \ref{sec:optimality}. The DSS problem is described and our result stated in Section \ref{sec:dss}, and our main theorem for the DSS problem is proved in Section \ref{sec:dss_proof}.


\section{Problem Formulation and Results}
\label{problem}
\subsection{Problem Formulation}
Let $N$ be a positive integer and define $[N] = \{1,2,\ldots,N\}$.
Let $x^n$ denote\footnote{When the length of the vector is particularly
important, we indicate it using a superscript.}
the source message in $\mathcal{X} ^n$, where $\mathcal{X} = [K]$ is the alphabet for the source.
We will call $n$ the \emph{blocklength} of the source.
We do not assume that a probability distribution over $\mathcal{X} ^n$ is given;
all of our results will be worst-case over this space. Given the
source sequence $x^n$, the encoder creates $N$ \emph{packets} 
(or \emph{messages}, or \emph{codewords}) via the functions
$$
f_\ell :\mathcal{X} ^n \mapsto \mathcal{X} ^{n R} \quad \ell \in \{1,\ldots,N\}.
$$
Note that we only
consider the problem in which all of the packets have the
same rate $R$. The encoder sends the packets
$$
(f_1(x^n),f_2(x^n), \ldots, f_N(x^n)),
$$
which we will often abbreviate as
$$
(C_1,C_2,\ldots,C_N).
$$
The decoder employs a function
$$
g : \prod_{\ell = 1}^N \mathcal{X}^{nR} \mapsto \{\mathcal{X} \cup {e}\}^n
$$
to reproduce the source given the received packets. The
fidelity of the reproduction is measured using the \emph{erasure
distortion measure}~\cite[p.~338]{cover2012elements}: for 
$x \in \mathcal{X}$ and
$\hat{x} \in \{\mathcal{X} \cup e\}$, define
\begin{equation}
\label{eq:erasure}
d(x,\hat{x}) =
\begin{cases}
0 &  \text{if $x = \hat{x}$} \\
1 &  \text{if $\hat{x} = e$} \\
\infty & \text{otherwise}.
\end{cases}
\end{equation}
We extend the single-letter distortion measure $d(\cdot,\cdot)$
to strings in the usual way
$$
d(x^n,\hat{x}^n) = \frac{1}{n}\sum_{i = 1}^n d(x_i,\hat{x}_i).
$$

We call the tuple $(f_1,\ldots,f_N,g)$ a \emph{code} for
the problem. We shall consider codes for which
the source $x^n$ can be perfectly reconstructed when
all of the packets are received unaltered, i.e.,
$$
\max_{x^n \in \mathcal{X}^n} d(x^n, g(C_\ell, \ell \in [N])) = 0.
$$
We call such codes \emph{feasible}.
For feasible codes, we shall consider how well the decoder
can reproduce the source when at most $T$ of the packets
are received in error
\begin{multline*}
D_T(f_1,\ldots,f_N,g) := \\
\max_{x^n \in \mathcal{X}^n} \max_{A \subseteq [N]: |A| \le T}
     \max_{\tilde{C}_A} d(x^n,g(C_{A^c}, \tilde{C}_A)).
\end{multline*}
Here $g(C_{A^c}, \tilde{C}_A)$ denotes the decoder's output
when its input is $C_\ell = f_\ell(x^n)$ for all $\ell \in A^c$ and
$\tilde{C}_\ell$ for all $\ell \in A$.\footnote{The problem can
be easily formulated using arbitrary distortion measures and
arbitrary distortion constraints, akin
to the general MD problem. But we shall focus 
exclusively on the problem as formulated here.}

\begin{defn}
The rate-distortion pair (R-D pair) $(R,D)$ is \emph{achievable} if for all $\epsilon >0$, there exists a feasible
code $(f_1,\ldots,f_N,g)$ for some blocklength with rate at most $R+\epsilon$ such that
$$
D_T(f_1,\ldots,f_N,g) \le D+ \epsilon.
$$
\end{defn}

\subsection{Main Result}

Our main result is the following.

\begin{theorem}
\label{the: Main}
Suppose the maximum number of altered packets $T$ satisfies
$T \ge 1$ and 
the number of packets $N$ satisfies 
$N \ge T + \lfloor \frac{T^2}{4} \rfloor + 2$.
\begin{enumerate}
\item If $0 \le R < \frac{1}{N-T}$, then there is no finite $D$ for
   which $(R,D)$ is achievable.\footnote{In a conference version of
   this result~\cite{fan2013polytope}, it was incorrectly asserted that 
   feasible codes do not exist if $0 \le R < \frac{1}{N-T}$. The correct
   statement is as given here.}
\item 
Let $F(T)$ denote $T + \lfloor \frac{T^2}{4} \rfloor + 1$. 
Then for any $\frac{1}{N-T} \le R \le \frac{1}{N-2T}$, the
rate-distortion pair
$$
\left(R,\frac{F(T)(N-T) (1 - (N - 2T) R)}{NT}\right)
$$
is achievable.
\end{enumerate}
\end{theorem}

The performance in part 2) is achieved using polytope codes
and should be compared against what can be obtained
using conventional MDS codes. Suppose we map $N - 2T$ source
symbols to $N$ coded symbols using an $(N,N - 2T)$ MDS code
(we can, if necessary, group several source symbols together
to ensure that the source alphabet is large enough to guarantee
the existence of such a code). Let each coded packet 
consist of exactly one of the coded symbols. The rate per packet
is then $R = 1/(N-2T)$, and since the minimum distance of the code is
$2T+1$~\cite{singleton1964maximum}, the decoder can always recover the 
source sequence exactly, 
even when there are $T$ errors. Thus this scheme achieves the
rate-distortion pair $(1/(N-2T),0)$.

\begin{figure}
\begin{center}
\includegraphics[clip=true,trim=.5in 2.5in 1in 2.5in,scale=.4]{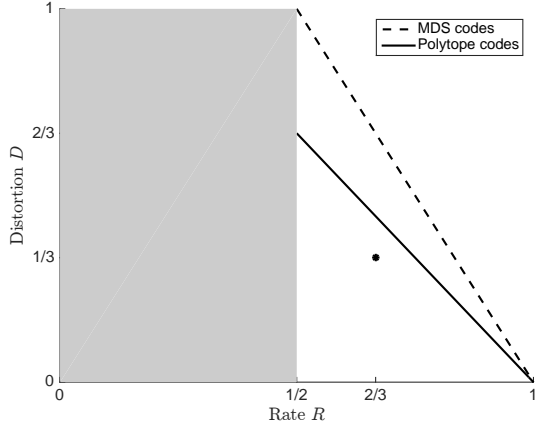}
\caption{Rate-distortion tradeoff for $N = 3$ packets and $T = 1$ error.
The dashed and solid lines indicate the achievable performance using 
MDS and polytope codes, respectively. The asterix indicates the rate-distortion
performance of the scheme in (\ref{eq:ebad3packet}). For rates below $1/2$,
finite distortion is unachievable for any feasible code.}
\label{fig:N3T1}
\end{center}
\end{figure}

\begin{figure}
\begin{center}
\includegraphics[clip=true,trim=.5in 2.5in 1in 2.5in,scale=.4]{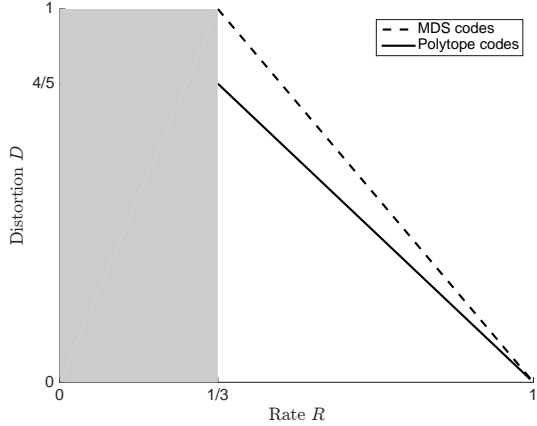}
\caption{Rate-distortion tradeoff for $N = 5$ packets and $T = 2$ errors.}
\label{fig:N5T2}
\end{center}
\end{figure}

On the other hand, if we use an $(N,N-T)$ MDS code, then the
decoder can reconstruct the source when there are no errors,
and since the minimum distance is $T+1$, it can always detect
when there are $T$ or fewer errors and output the all-erasure 
string in response. Hence
this code can achieve the rate-distortion pair $(1/(N-T),1)$.
A simple time-sharing argument shows that the line connecting
these points
$$
\left(R,\frac{N-T}{T} - \frac{(N-T)(N - 2T)}{T} R\right)
$$
is achievable. This is shown in Fig.~\ref{fig:N3T1} for $N = 3$
and $T = 1$ and in Fig.~\ref{fig:N5T2} for $N = 5$ and $T = 2$,
along with the achievable rate-distortion pairs from
Theorem~\ref{the: Main}. We see that Theorem \ref{the: Main}
does strictly better.

When $N = 3$ and $T = 1$, there is actually a simple design that
is not dominated by the above schemes.
When $R= \frac{2}{3}$, let the blocklength of the source message be three
and write the source as $(x_1,x_2,x_3)$. We transmit 
\begin{align}
\label{eq:ebad3packet}
 (x_1,x_2)  \quad (x_2,x_3) \quad (x_3,x_1)
\end{align}
as the three packets.
The decoder can check whether the copy of $x_i$ is the same
between the two packets in which it appears for each $i$. 
If the two packets 
have the same value of $x_i$, then this common value must be correct.
Since the channel can alter at most one packet, there can be at most
two components of $(x_1,x_2,x_3)$ on which there is disagreement. If there 
is disagreement about two source components, however, then the decoder
can identify which packet was altered, exclude it, and then
determine all of the source components from the remaining packets.
Thus the maximum number of components  about which the
decoder can be uncertain is one. It follows that the R-D pair
$(2/3,1/3)$ is achievable. This point lies outside the region
achieved by polytope codes, as shown in Fig.~\ref{fig:N3T1}.

Since the rate-distortion pair $(1/(N-2T),0)$ is achievable, and
the set of achievable pairs is convex, to show part 2) of 
Theorem~\ref{the: Main} it suffices to show that 
$$
\left(\frac{1}{N-T},\frac{F(T)}{N}\right)
$$
is achievable. In the next section, we will show how polytope 
codes can be used toward this end. Note that, per the statement
of Theorem~\ref{the: Main}, the resulting scheme can only be applied
when $N \ge F(T) + 1$. In particular, the blocklength must grow
with the square of the number of errors. This is undesirable;
one would prefer to have linear scaling. In Section~\ref{sec:optimality},
we show that this quadratic scaling cannot be improved by
changing the decoder---it is intrinsic to the code itself.
Of course, since $N$ represents the number of independent
paths in the network between the encoder and the decoder,
we are generally interested in small values of $N$ and $T$,
so that the scaling behavior is not paramount.

\section{Polytope Codes}
\label{sec:Polytope Codes}
Polytope codes were introduced by Kosut, Tong, and Tse~\cite{KosutTongTse:14IT} in
the context of network coding with adversarial nodes. Polytope codes
are akin to linear MDS codes, except that the arithmetic operations
are performed over the reals and extra low rate ``check'' information 
is included in the transmission. Our construction is
somewhat simpler than the one given in~\cite{KosutTongTse:14IT}. To understand
this construction it is helpful to begin with the special case
in which there are $N = 3$ packets subject to at most $T = 1$
error.

\subsection{$N = 3$, $T = 1$ case}

One trivial design for this case is to simply send the true source 
sequence in all three packets. Since there is at most one error, the decoder
can always recover the source sequence by using a majority rule. 
That is, it can recover the source exactly when there are no errors
but also when there is one. As such, this scheme achieves the 
rate-distortion pair $(1,0)$. This scheme is unsatisfactory, however,
since it is wasteful when there no errors.

One may consider using a $(3,2)$ MDS code instead. For instance,
we could choose the blocklength $n = 2$ and encode two source
symbols $x_1$ and $x_2$ into three packets as
\begin{equation}
\label{eq:MDS}
x_1 \quad \quad x_2 \quad \quad x_1 \oplus x_2,
\end{equation}
where $\oplus$ denotes modulo arithmetic. The decoder can
determine whether a single error has been introduced by verifying
whether the received packets satisfy the linear relation in
(\ref{eq:MDS}). If so, then there are no errors, and the decoder
can reproduce the source exactly. Thus it is feasible. If not,
then the decoder knows that one error is present, but it has
no way of identifying which packet is in error. Since there
is an infinite penalty for guessing a source symbol incorrectly,
it must output the all-erasure string, achieving the
rate-distortion pair $(1/2,1)$.
The striking thing about this example is that the decoder
always receives at least one of the two source symbols
correctly; the problem is that it does not know which of
the two is correct.

Now suppose that the source is viewed as a pair of
vectors of positive integers of length $N_0$, $x_1^{N_0}$
and $x_2^{N_0}$, and the three transmitted packets consist of 
\begin{equation}
\label{eq:poly3}
x_1^{N_0} \quad \quad x_2^{N_0} \quad \quad x_1^{N_0} + x_2^{N_0},
\end{equation}
where now the addition is performed over the reals. We
also send the quantities
\begin{equation}
\label{eq:innerprods3}
\langle x_i^{N_0}, x_j^{N_0} \rangle 
\end{equation}
for all $i$ and $j$ as part of each packet. As before,
the decoder can always detect whether an error has been 
introduced. If it detects no error, it can output the
source sequence correctly. But now if it detects an error,
it can always identify at least one of the three packets as
correct, by the following reasoning. Since the inner
products in (\ref{eq:innerprods3}) are included in all
three packets, they can always be recovered correctly.
Let 
\begin{equation}
\label{eq:poly3tilde}
\tilde{x}_1^{N_0} \quad \quad \tilde{x}_2^{N_0} \quad  \quad
       \tilde{x}_3^{N_0},
\end{equation}
denote the vectors in the three received packets,
and assume that exactly one of them has been
altered. If for any $i$ we have
$$
||\tilde{x}_i^{N_0}||^2 \ne ||x_i^{N_0}||^2,
$$
then we know that the $i$th packet is in error 
and the other two must be correct. So we shall
assume that
$$
||\tilde{x}_i^{N_0}||^2 = ||x_i^{N_0}||^2,
$$
for all $i$.

Now construct a graph with nodes 
$\tilde{x}_1^{N_0}$, $\tilde{x}_2^{N_0}$,
and $\tilde{x}_3^{N_0}$ and
an edge between $\tilde{x}_i^{N_0}$ and
$\tilde{x}_j^{N_0}$ (for $i \ne j$) if
$$
\langle \tilde{x}_i^{N_0}, \tilde{x}_j^{N_0} \rangle
   = \langle {x}_i^{N_0}, {x}_j^{N_0} \rangle
$$
We call this the \emph{syndrome graph}. 
Consider the number of edges in the syndrome graph.
If the syndrome graph is fully connected, then for some
collection of constants $a_{ij}$  we must have
\begin{align}
\label{eq:3innerproductstart}
||\tilde{x}_3^{N_0} - \tilde{x}_1^{N_0} - \tilde{x}_2^{N_0}||^2
  & = \sum_{i,j} a_{ij} \langle \tilde{x}_i^{N_0}, \tilde{x}_j^{N_0} \rangle \\
  & = \sum_{i,j} a_{ij} \langle x_i^{N_0}, x_j^{N_0} \rangle \\
  & = ||x_3^{N_0} - x_1^{N_0} - x_2^{N_0}||^2  \\
\label{eq:3innerproductend}
  & = 0.
\end{align}
Thus 
$$
\tilde{x}_3^{N_0} = \tilde{x}_1^{N_0} + \tilde{x}_2^{N_0},
$$
which contradicts the assumption that one of the these vectors
was altered. 

Thus the graph must be missing at least one edge. Since only
one packet can be received in error, the graph cannot be missing
all three edges, however. Thus it must have either one edge or two.
If it has exactly one edge, then the vector with no edges must be
the one in error, so the other two vectors can be identified as
correct. If the graph has two edges, then the vector with two
edges must be correct. In the end, then, the decoder can always
recover at least one of the transmitted packets correctly. This
is of course not the same as recovering one of the source vectors---if
the decoder recovers $x_3^{N_0}$ then it cannot reproduce any of the
source symbols with certainty. But using a ``layering'' argument
one can transform this code into one for which decoding any of
the three transmitted packets correctly allows one to recover
some positive fraction of the source symbols correctly
(see Section~\ref{sec:proof of theorem}).

The property that the decoder can always correctly recover 
a transmitted packet even when the number of errors is outside
the decoding radius of the code we call \emph{guaranteed partial
decodability.} This property comes at slight cost in rate 
compared with conventional MDS codes; one must send the norms
and inner products in (\ref{eq:innerprods3}) in addition to the vectors,
and $x_3^{N_0}$ can take larger values than either $x_1^{N_0}$
or $x_2^{N_0}$ because the addition in (\ref{eq:poly3}) is done over
the reals. But in the limit of a large source blocklength, this 
penalty can be made arbitrarily small, and the rate can be
made arbitrarily close to $1/2$.

We next describe how to extend this idea to general $N$ and
$T$. The resulting construction is then used to prove
Theorem~\ref{the: Main}. See~\cite{fan2013polytope} for a slightly different 
decoding algorithm that yields the same performance.

\subsection{General $(N,T)$: Source}
Consider a source message $x^n$ $(x^n \in \mathcal{X} ^n)$ with length $n=(N-T)N_{0} K_0$ for some large natural numbers $N_0$ and $K_0$. Divide the message into $(N-T)N_0$ subvectors, each having $K_0$ symbols. We can use a $K_0$-length vector (each entry taken from $[K]$) to represent $K^{K_0}$ integers $\{1,...,K^{K_0}\}$; here we use $(0,...,0)$ to represent $K^{K_0}$. Thus, the original source message can also be viewed as an integer vector with length $(N-T)N_0$. Moreover, $x^n$ can be viewed as a concatenation of $N-T$ vectors, each having $N_0$ entries in $\{1,...,K^{K_0}\}$. In what follows, we will
view the source vector in this way and write
\[x^n = (x_{1,K_0}^{N_0},...,x_{N-T,K_0}^{N_0}).\]
\subsection{Encoding Functions}

The encoding is performed with the aid of an eligible generator
matrix. 
\begin{defn}
\label{defn: eligible matrix}
$A$ is an \textit{eligible $(N,N-T)$-generator matrix} if 
its entries are nonnegative integers and
\begin{enumerate}
\item $A$ is an $N \times (N-T)$ matrix of the following form:
\[ A=\left[ {\begin{array}{*{20}{c}}
   1 & 0 &  \cdots  & 0  \\
   0 & 1 &  \ddots  &  \vdots   \\
    \vdots  &  \ddots  &  \ddots  & 0  \\
   0 &  \cdots  & 0 & 1  \\
   {a _{1,1}} & {a _{1,2}} &  \cdots  & {a _{1,N-T}}  \\
    \vdots  &  \vdots  &  \vdots  &  \vdots   \\
   {a _{T,1}} & {a _{T,2}} &  \cdots  & {a _{T,{N-T}}}  \\
\end{array}} \right],\] 

\item Every $(N-T) \times (N-T)$ submatrix of $A$ is nonsingular.
\end{enumerate}
\end{defn}
The existence of such matrix is guaranteed by the following lemma.
\begin{lemma}
\label{lemma:generator}
For any $T \ge 1$ and $N \ge T$ there exists an eligible $(N,N-T)$-generator
matrix of the form
\begin{equation}
\label{eq:Alemma}
A=\left[ {\begin{array}{*{20}{c}}
   1 & 0 &  \cdots  & 0  \\
   0 & 1 &  \ddots  &  \vdots   \\
    \vdots  &  \ddots  &  \ddots  & 0  \\
   0 &  \cdots  & 0 & 1  \\
   {\alpha _1^1} & {\alpha _1^2} &  \cdots  & {\alpha _1^{N - T}}  \\
    \vdots  &  \vdots  &  \vdots  &  \vdots   \\
   {\alpha _T^1} & {\alpha _T^2} &  \cdots  & {\alpha _T^{N - T}}  \\
\end{array}} \right],
\end{equation}
where $\alpha_1, \ldots, \alpha_T$ are distinct positive integers.
We call such a matrix a $\emph{V-matrix}$,
since its lower portion has a Vandermonde structure.
\end{lemma}

\begin{IEEEproof}
We find the required $\alpha_1, \ldots, \alpha_T$ by induction.
Clearly there exists a positive integer $\alpha_1$ such that
$$
 A_1 =\left[ {\begin{array}{*{20}{c}}
   1 & 0 &  \cdots  & 0  \\
   0 & 1 &  \ddots  &  \vdots   \\
    \vdots  &  \ddots  &  \ddots  & 0  \\
   0 &  \cdots  & 0 & 1  \\
   {\alpha _1^1} & {\alpha _1^2} &  \cdots  & {\alpha _1^{N - T}}  \\
\end{array}} \right],
$$
is such that every $(N-T)\times(N-T)$ submatrix is nonsingular.
Indeed, taking $\alpha_1 = 1$ suffices. Now suppose we have positive integers
$\alpha_1, \ldots \alpha_{t-1}$ such that every $(N-T)\times(N-T)$ submatrix
of 
$$
A_{t-1}=\left[ {\begin{array}{*{20}{c}}
   1 & 0 &  \cdots  & 0  \\
   0 & 1 &  \ddots  &  \vdots   \\
    \vdots  &  \ddots  &  \ddots  & 0  \\
   0 &  \cdots  & 0 & 1  \\
   {\alpha _1^1} & {\alpha _1^2} &  \cdots  & {\alpha _1^{N - T}}  \\
    \vdots  &  \vdots  &  \vdots  &  \vdots   \\
   {\alpha _{t-1}^1} & {\alpha _{t-1}^2} &  \cdots  & 
             {\alpha _{t-1}^{N - T}}  \\
\end{array}} \right]
$$
is nonsingular. Consider the matrix
$$
A_{t}=\left[ {\begin{array}{*{20}{c}}
   1 & 0 &  \cdots  & 0  \\
   0 & 1 &  \ddots  &  \vdots   \\
    \vdots  &  \ddots  &  \ddots  & 0  \\
   0 &  \cdots  & 0 & 1  \\
   {\alpha _1^1} & {\alpha _1^2} &  \cdots  & {\alpha _1^{N - T}}  \\
    \vdots  &  \vdots  &  \vdots  &  \vdots   \\
   {\alpha _{t}^1} & {\alpha _{t}^2} &  \cdots  & 
             {\alpha _{t}^{N - T}}  \\
\end{array}} \right],
$$
viewed as a function of the variable $\alpha_t$. For any 
given $(N-T)\times(N-T)$
submatrix of $A_t$ of the form
\begin{equation}
\label{eq:atfunc}
 \left[ {\begin{array}{*{20}{c}}
   \multicolumn{4}{c}{\tilde{A}} \\
   {\alpha _{t}^1} & {\alpha _{t}^2} &  \cdots  & 
             {\alpha _{t}^{N - T}}  \\
\end{array}} \right],
\end{equation}
there must exist a natural number $\alpha_t$ such that this particular
$(N-T) \times (N-T)$ matrix is nonsingular, by the following reasoning.
The rows of $\tilde{A}$ are linearly independent by the induction
hypothesis. Let $[v_1 \ \ v_2 \ \ \cdots v_{N-T}]$ be a nonzero row vector
such that
\begin{equation}
\left[ {\begin{array}{*{20}{c}}
   \multicolumn{4}{c}{\tilde{A}} \\
   {v_1} & {v_2} &  \cdots  & 
             {v_{N - T}}  \\
\end{array}} \right],
\end{equation}
is full rank. Then let $[\tilde{v}_1 \ \ \tilde{v}_2 \ \ \cdots 
   \tilde{v}_{N-T}]$ denote the component of $[v_1 \ \ v_2 \ \ \cdots v_{N-T}]$
that is orthogonal to the row space of $\tilde{A}$ and
note that $[\tilde{v}_1 \ \ \tilde{v}_2 \ \ \cdots 
   \tilde{v}_{N-T}]$  must be nonzero.
Then we can find a natural number $\alpha_t$ so that
$$
\sum_{i = 1}^{N-T} \tilde{v}_i \alpha_t^i \ne 0.
$$
This follows from the fact that the left-hand side is a nonzero
$(N-T)$-degree polynomial in $\alpha_t$, so that there must be 
a positive integer that is not a root. We conclude that the 
determinant of the $(N-T)\times(N-T)$ matrix in (\ref{eq:atfunc}),
which is evidently an $(N-T)$-degree polynomial in $\alpha_t$,
is not identically zero. 

Next we show that there is one choice of $\alpha_t$ that ensures
that every $(N-T)\times(N-T)$ submatrix of $A_t$ is nonsingular.
The determinant of any given $(N-T)\times(N-T)$ submatrix is
a nonzero $(N-T)$-degree polynominal in $\alpha_t$, as noted 
earlier. Thus it has at most $(N-T)$ roots according to fundamental 
theorem of algebra. Thus
all of the submatrices together have at most ${N - T + t -1
\choose N - T - 1} (N-T)$ roots. Since this is finite, there
must exist a natural number $\alpha_t$ that is not a root of
any of these polynomials.
\end{IEEEproof}
The encoding functions are then as follows:
\begin{enumerate}
\item We generate $N$ vectors, $y_1^{N_0} \ldots
y_N^{N_0}$ via the linear transformation
$$
\left[
\begin{array}{c}
y_{1,K_0}^{N_0} \\
\vdots \\
y_{N,K_0}^{N_0}
\end{array}
\right]
  = A 
\left[
\begin{array}{c}
x_{1,K_0}^{N_0} \\
\vdots \\
x_{N-T,K_0}^{N_0}
\end{array}
\right],
$$
where $A$ is an eligible $(N,N-T)$-generator matrix provided by 
Lemma~\ref{lemma:generator}. In particular, we have
$$
y_{i,K_0}^{N_0} = x_{i,K_0}^{N_0}
$$
for all $1 \le i \le N-T$. We assume that each vector
is encoded using $({K_0} + \left\lceil {{{\log }_K} (\alpha (N-T))} 
 \right\rceil)N_0$ symbols, where $\alpha = \max_{i,j} \alpha_{i,j}.$

\item We also transmit $(N-T) + {N - T \choose 2}$ norms/inner products: \[F_{ij}= \langle x_{i,K_0}^{N_0},x_{j,K_0}^{N_0}\rangle,\forall 1\le i \le j\le N-T\] in all $N$ packets. This requires that $ \lceil (2 K_0+\log_K N_0)[(N-T) + {N - T \choose 2}]$ extra symbols to be included in each packet. 
\end{enumerate}

\subsection{General $(N,T)$: Decoding Functions}
\label{decoding}
The decoder receives $\bar {y}_{1,K_0}^{N_0},...,\bar {y}_{N,K_0}^{N_0}$ and the 
norms/inner products between $\{x_1^{N_0},...,x_{N-T}^{N_0}\}$.
The decoder will identify a subset of the components of 
$y_1^{N_0},...,y_N^{N_0}$ that it is sure have been 
unaltered.\footnote{Later we will show how to use this 
identification to prove Theorem~\ref{the: Main}.}
We first note that the norms and inner products can always be recovered
without error.

\begin{lemma}
\label{lem:inner}
The decoder can correctly recover $F_{ij}$ for $i,j \in \{1,...,N-T\}$ when $N \ge 2T+1$. Since $y_1^{N_0},...,y_N^{N_0}$ are linear combinations of $x_1^{N_0},..,x_{N-T}^{N_0}$. This means that we can correctly recover $F_{ij}=\langle y_i^{N_0},y_j^{N_0} \rangle$ for 
$i,j \in \{1,...,N\}$.
\end{lemma}
The proof of this lemma is straightforward and omitted. 

Use a graph $G$ with $N$ vertices $V=\{v_1,v_2,...,v_N\}$ to represent the $N$ received packets. The $i$th received packet is $\bar {C}_i$, which is composed of the $K$-symbol representations of $\bar{y}_i^{N_0}$ and $\bar {F}_{j_1j_2}^{(i)}$ ($1 \le j_1 \le j_2\le N-T$). According to Lemma \ref{lem:inner}, we can correctly recover $F_{ij}=\langle y_i^{N_0}, y_j^{N_0} \rangle$. We draw an edge between vertex $v_i$ and vertex $v_j$ $(i\ne j)$ iff 
\[\langle \bar {y}_i^{N_0},\bar {y}_j^{N_0} \rangle = F_{ij}.\]
We draw a self-loop on vertex $v_i$ iff 
\[ \langle \bar {y}_i^{N_0},\bar {y}_i^{N_0} \rangle = F_{ij}.\]
As in the $N = 3, T = 1$ case, 
we call this the \emph{syndrome graph.}

The decoder then performs the following operations:
\begin{enumerate}
\item Delete all vertices with no loops and their incident edges in the syndrome graph. Let $\hat G = (\hat{V},\hat{\mathcal{E}})$ denote the new graph.
\item Let $V'$ be the set of vertices $v_i$ in $\hat{V}$ such that $v_i$ is contained in a clique of size at least $N-T$ in $\hat G$.
\item Let $V^*$ be the set of vertices $v_i$ in $V'$ such that $(v_i,v_j) \in \hat{\mathcal{E}}$ for all $v_j$ in $V'$.
\item Output the codewords corresponding to the vertices in $V^*$ as
correct.
\end{enumerate}

We shall show that
the rate of this code can be made arbitrarily close to $1/(N-T)$.
We shall then prove that the codewords $\bar y_i^{N_0}$ on channels 
corresponding to the vertices $v_i \in V^*$ are correct.

\subsection{General $(N,T)$: Coding Rate}
\label{coding rate}
\begin{proposition}
For any $\epsilon > 0$, there exists natural numbers $K_0$ and $N_0$
such that the rate of each packet does not exceed $1/(N-T) + \epsilon$.
\end{proposition}

\textit{Proof:}
The rate of each packet is upper bounded by
\begin{align}
\nonumber
& \frac{{({K_0} + \left\lceil {{\log_K}(\alpha (N-T))} \right\rceil ){N_0}}}{{{K_0}{N_0}(N - T)}} \\
\label{eq:rateformula}
& +\frac{\lceil2K_0 + \log_K N_0\rceil\left((N-T) 
    + {N - T \choose 2}\right)}
    {{{K_0}{N_0}(N - T)}},
\end{align}
where we recall that $\alpha = \max_{i,j} \alpha_{i,j}$.
If we let $N_0 = K_0$ and send both to infinity,
the second term tends to zero while the first term tends to 
$1/(N-T)$.

\subsection{General $(N,T)$: Partial Decodability of Polytope Codes}

We are interested in polytope codes because of the following
property.

\begin{theorem}
\label{the: SinglePoly}
Given $T$, when $N \ge T + \left\lfloor {\frac{{{T^2}}}{4}} \right\rfloor+ 2$, the decoder can identify least $N-T-\left\lfloor \frac{T^2}{4}\right\rfloor-1$ of the transmitted packets as being received correctly.
\end{theorem}

We shall prove Theorem~\ref{the: SinglePoly} via a sequence of lemmas. The first
two establish that the codewords associated with nodes in $V^*$
were received correctly.

\begin{lemma}
\label{lem:noloop}
Suppose the $k$ packets $i_1,\ldots,i_k$ are unaltered, and let 
$i_{k+1}$ be some other packet for which there exists $l_1,\ldots,l_k$ 
such that
\begin{equation}\label{eq:linear_constraint}
y_{i_{k+1}}^{N_0}=\sum_{j=1}^k l_j y_{i_j}^{N_0}.
\end{equation}
If there is a self-loop on $v_{i_k+1}$ in $G$, and $(v_{i_{k+1}},v_{i_j})\in\mathcal{E}$ for all $j\in\{1,\ldots,k\}$, then the codeword $\bar{y}_{i_{k+1}}^{N_0}$ in packet $i_{k+1}$ is also unaltered.
\end{lemma}
\begin{IEEEproof}
We may rewrite \eqref{eq:linear_constraint} as
\begin{equation}\label{eq:quadratic_constraint}
\Big\| y_{i_{k+1}}^{N_0} - \sum_{j=1}^k l_j y_{i_j}^{N_0}\Big\|^2=0.
\end{equation}
Since there is a self-loop on $v_{i_{k+1}}$, 
\[
\langle \bar{y}_{i_{k+1}}^{N_0}, \bar{y}_{i_{k+1}}^{N_0}\rangle = 
\langle {y}_{i_{k+1}}^{N_0}, {y}_{i_{k+1}}^{N_0}\rangle.
\]
Moreover, since there is an edge $(v_{i_{k+1}},v_{i_j})$ for all $j\in\{1,\ldots,k\}$, 
\[
\langle \bar{y}_{i_{k+1}}^{N_0},\bar{y}_{i_j}^{N_0}\rangle= 
\langle {y}_{i_{k+1}}^{N_0},{y}_{i_j}^{N_0}\rangle.
\]
By expanding the left-hand side of \eqref{eq:quadratic_constraint}
in terms of inner products, as in 
(\ref{eq:3innerproductstart})-(\ref{eq:3innerproductend}), 
we have that
\begin{align*}
0
&=\Big\| y_{i_{k+1}}^{N_0} - \sum_{j=1}^k l_j y_{i_j}^{N_0}\Big\|^2
\\&= \Big\| \bar{y}_{i_{k+1}}^{N_0} - \sum_{j=1}^k l_j y_{i_j}^{N_0}\Big\|^2
\\&= \Big\| \bar{y}_{i_{k+1}}^{N_0} - y_{i_{k+1}}^{N_0}\Big\|^2
\end{align*}
where we have used the assumption that packets $i_1,\ldots,i_k$ are unaltered, and \eqref{eq:linear_constraint}. This proves that packet $i_{k+1}$ is unaltered.
\end{IEEEproof}

\begin{lemma}
\label{lemma:claim1}
For any $i\in V^*$, we have $\bar{y}_i^{N_0}=y_i^{N_0}$.
\end{lemma}
\begin{IEEEproof}
There must exist $N-T$ packets that are unaltered. Suppose they are packets $i_1,\ldots,i_{N-T}$. Then $v_{i_1},\ldots,v_{i_{N-T}}$ must form a clique in the syndrome graph $\hat{G}$. From the definition of $V^*$, for any vertex $i\in V^*$, there is a self-loop on $i$ and $(i,v_{i_j})\in\mathcal{E}$ for all $j\in\{1,\ldots,N-T\}$. By construction, every $(N-T)\times (N-T)$ submatrix of generator matrix $A$ is nonsingular. This implies that the vector $y_i^{N_0}$ can be represented as a linear combination of the other $N-T$ vectors
\[
y_{i_{k+1}}^{N_0}=\sum_{j=1}^{N-T} l_j y_{i_j}^{N_0}
\]
for some linear coefficients $l_j$. By Lemma~\ref{lem:noloop}, the codeword $y_i^{N_0}$ in packet $i$ is unaltered.
\end{IEEEproof}

The final lemma lower bounds the size of $V^*$. It is a purely
graph-theoretic assertion that may have independent uses.

\begin{lemma}
\label{lemma:claim2}
Consider an undirected graph $\mathcal{G} = (\mathcal{V},
\mathcal{E})$ with at least
$N - T$ nodes in which every node has a self-loop. Let $V'$
denote the set of nodes that are contained in a clique of size
at least $N - T$, and suppose that $V'$ is not empty. Let
$$
V^* = \{v \in V': (v,\tilde{v}) \in \mathcal{E} \ \forall
   \tilde{v} \in V'\}.
$$
Then we have $|V^*| \ge N-F(T)$, where $F(T)$ is defined in 
Theorem~\ref{the: Main}.
\end{lemma}

\begin{IEEEproof}
For any set of edges $\mathcal{E}_0$, let
\[\mathcal{N}(v_i,\mathcal{E}_0) :=\{v_j \in V' \backslash \{v_i\}: (v_i,v_j) \notin \mathcal{E}_0\},\]
We construct a set of edges $\mathcal{E}' \supset \mathcal{E}$ as follows. 
Begin by setting $\mathcal{E}' = \mathcal{E}$. If there is a pair $v_i,v_j \in V'$ such that $(v_i, v_j) \notin \mathcal{E}'$ and 
\[|\mathcal{N} (v_i, \mathcal{E}')| >1, |\mathcal{N} (v_j, \mathcal{E}')| >1,\]
then add $(v_i,v_j)$ to $\mathcal{E}'$. Repeat until there is no such pair $v_i, v_j$. Note that for the resulting $\mathcal{E}'$, for 
$v_i \in V'$, $\mathcal{N}(v_i, \mathcal{E}') = 0$ if and only if $\mathcal{N} (v_i, \mathcal{E})=0$. Thus
\[V^*=\{v_i \in V': \mathcal{N} (v_i, \mathcal{E}')=0\}.\]
Moreover, for any pair $(v_i,v_j)\in V'$ with $(v_i,v_j) \notin \mathcal{E}'$, either $|\mathcal{N} (v_i, \mathcal{E}')|=1$ or $|\mathcal{N} (v_j, \mathcal{E}')|=1$. For convenience, we write $\mathcal{N}(v_i) := \mathcal{N}(v_i, \mathcal{E}')$ from now on.

Let $v_{i_0}$ be an element of $V'$ maximizing $|\mathcal{N}(v)|$, and let 
\[l_0 := |\mathcal{N}(v_{i_0})|.\]
Each element $v_i \in V'$ is contained in a clique of $\mathcal{C}(v_i)$ of size exactly $N-T$.\footnote{There may be several such cliques, in which case
$\mathcal{C}(v_i)$ can be chosen to be any one of them.}
 Since $\mathcal{E}' \supset \mathcal{E}$, 
$\mathcal{C}(v_i)$ is also a clique on the graph with edges $\mathcal{E}'$. Let $\mathcal{C}_0 = \mathcal{C}{v_{i_0}} \backslash \{v_{i_0}\}$.
Fix $v_{i_1} \in \mathcal{C}_0$, and suppose $(v_{i_1},v_l) \notin \mathcal{E}'$ for $v_l \in V'$. We claim that $v_l$ cannot be in $\mathcal{N}(v_{i_0})$. If it were, then $\mathcal{N} (v_l) \ge 2$, in which case $l_0 \ge 2$, which would imply that $|\mathcal{N}(v_{i_0})| \ge 2$. But 
$(v_{i_0}, v_l) \notin \mathcal{E}'$, which contradicts the construction of $\mathcal{E}'$. Moreover, $v_l$ cannot be in $\mathcal{C}(v_{i_0})$ 
by definition. Hence, if $(v_{i_1}, v_l) \ne \mathcal{E}'$, then $v_l \in \mathcal{D}$, where 
\[\mathcal{D}:= V' \backslash \mathcal{N}(v_{i_0}) \backslash \mathcal{C}(v_{i_0}).\]
In particular, if $v_{j} \in \mathcal{C}_0 \cap V'\backslash V^*$, then $(v_j,v_k) \notin \mathcal{E}'$ 
for some $v_k \in \mathcal{D}$; i.e. $v_j \in \mathcal{N}(v_k)$. Thus
\begin{align*}
V' \backslash V^* & \subset (V' \backslash \mathcal{C}_0 \backslash V^*)\cup (V' \cap \mathcal{C}_0 \backslash V^*) \\
&\subset \{v_{i_0}\} \cup (V' \backslash \mathcal{C}(v_{i_0})) \cup \cup_{v \in \mathcal{D}}(\mathcal{N}(v) \cap \mathcal{C}_0)\\
&\subset  \{v_{i_0}\} \cup \mathcal{N}(v_{i_0}) \cup \mathcal{D} \cup \cup_{v \in \mathcal{D}}(\mathcal{N}(v) \cap \mathcal{C}_0).
\end{align*}
Hence,
\begin{align}
\nonumber
|V'|-|V^*| &\le 1+ |\mathcal{N}(v_{i_0})|+|\mathcal{D}| +\Sigma_{v\in \mathcal{D}}|\mathcal{N}(v)|\\
\label{eq:Vstarbound}
&\le (|\mathcal{D}|+1)(l_0+1),
\end{align}
where we have used the fact that $|\mathcal{N}(v)| \le l_0$ for all $v\in V'$. Since $\mathcal{N}(v_{i_0})$, $\mathcal{C}(v_{i_0}) \subset V'$ 
and $\mathcal{N}(v_{i_0}) \cap \mathcal{C}(v_{i_0}) = \emptyset$,
\[|\mathcal{D}|= |V'|- |\mathcal{N}(v_{i_0})|-|\mathcal{C}(v_{i_0})| = |V'|-l_0+T-N.\]
Substituting this into (\ref{eq:Vstarbound}) gives
\begin{align*}
|V^*| &\ge |V'|- (|\mathcal{D}|+1)(l_0+1) \\
&= |V'|-(T-l_0+|V'|-N+1)((l_0+1) \\
&\ge N- (T-l_0+1)(l_0+1)\\
&\ge N- F(T).
\end{align*}
\end{IEEEproof}

\begin{IEEEproof}[Proof of Theorem~\ref{the: SinglePoly}]
For each $i \in V^*$, we have $\bar{y}_i^{N_0} = y_i^{N_0}$ by 
Lemma~\ref{lemma:claim1} and $|V^*| \ge N - F(T)$ by Lemma~\ref{lemma:claim2}.
\end{IEEEproof}

\section{Proof of Theorem \ref{the: Main}}
\label{sec:proof of theorem}
We next show how to use polytope codes to create a code for our
original problem. The main difficulty is that, in a polytope
code, some of the packets contain only parities, and even if
the decoder can determine such packets with certainty, it cannot
necessarily recover any of the original source symbols. 
We circumvent this issue with a layered construction. First
we prove the impossibility result in part 1).

\subsection{Proof of Theorem \ref{the: Main} Part 1)}

Fix $0 \le R < \frac{1}{N - T}$ and $\epsilon > 0$ such that $R + \epsilon
< \frac{1}{N - T}$. If there does not exist a feasible code with rate at
most $R + \epsilon$ then the conclusion is immediate. Otherwise,
consider any feasible code with rate at most $R + \epsilon$, and 
let $n$ denote the length of the source string that it encodes.

Consider endowing the space $\mathcal{X}^n$ with an i.i.d. uniform
probability distribution. Since the code is feasible, the source
string must be a function of the messages, i.e.
$$
H(x^n|C_1,\ldots,C_N) = 0.
$$
Since $C_1, \ldots, C_N$ are also deterministic functions
of the source string, we must have
$$
H(C_1,\ldots,C_N) = H(x^n) = n \log K.
$$
Therefore
\begin{align*}
& H(C_1,\ldots,C_T|C_{T+1},\ldots, C_N) \\
  & \ge H(C_1,\ldots,C_N) - H(C_{T+1},\ldots,C_N) \\
  & \ge H(C_1,\ldots,C_N) - \sum_{i = T+1}^N H(C_i) \\
  & = n \log K - \sum_{i = T+1}^N H(C_i) \\
  & \ge n \log K - (N - T) n (R+\epsilon) \log K \\
  & > 0.
\end{align*}
Thus $(C_1,\ldots,C_T)$ is not a deterministic function
of $(C_{T+1},\ldots,C_N)$. It follows that there must exist
two source sequences $x_1^n$ and $x_2^n$ such that
$x_1^n \ne x_2^n$,
\begin{align*}
f_i(x_1^n) & \ne f_i(x_2^n) \quad \text{for some $1 \le i \le T$} \\
\intertext{and}
f_j(x_1^n) & = f_j(x_2^n) \quad \text{for all $T+1 \le j \le N$}.
\end{align*}

Since the code is feasible, when the decoder receives the message
$$
(f_1(x_1^n), f_2(x_1^n), \ldots, f_N(x_1^n)),
$$
it must output string $x_1^n$. But then the decoder will also output
$x_1^n$ if the true source sequence is $x_2^n$ and the adversary
alters the first $T$ packets so that
\begin{align*}
(f_1(x_1^n), \ldots, f_T(x_1^n), f_{T+1}(x_2^n), \ldots, f_N(x_2^n)) \\
= (f_1(x_1^n), \ldots, f_T(x_1^n), f_{T+1}(x_1^n), \ldots, f_N(x_1^n))
\end{align*}
is received. Since $x_1^n$ and $x_2^n$ are different, the distortion
of the code is infinite.

\subsection{Proof of Theorem \ref{the: Main} Part 2)}

As noted earlier it suffices to show that the R-D pair
$(\frac{1}{N-T},\frac{F(T)}{N})$ is achievable. To show this
we use a ``layered'' construction in which we use $N$ polytope codes
whose transformation matrices are row rotations of each other. Divide
the source into $N$ equal-sized parts. The first part is encoded into
packets using a polytope code with transformation matrix
 \[ A=\left[ {\begin{array}{*{20}{c}}
   1 & 0 &  \cdots  & 0  \\
   0 & 1 &  \ddots  &  \vdots   \\
    \vdots  &  \ddots  &  \ddots  & 0  \\
   0 &  \cdots  & 0 & 1  \\
   {\alpha _1^1} & {\alpha _1^2} &  \cdots  & {\alpha _1^{N - T}}  \\
    \vdots  &  \vdots  &  \vdots  &  \vdots   \\
   {\alpha _T^1} & {\alpha _T^2} &  \cdots  & {\alpha _T^{N - T}}  
\end{array}} \right].\]
The second part is encoded using the transformation matrix
 \[ A=\left[ {\begin{array}{*{20}{c}}
   {\alpha _T^1} & {\alpha _T^2} &  \cdots  & {\alpha _T^{N - T}}  \\
   1 & 0 &  \cdots  & 0  \\
   0 & 1 &  \ddots  &  \vdots   \\
    \vdots  &  \ddots  &  \ddots  & 0  \\
   0 &  \cdots  & 0 & 1  \\
   {\alpha _1^1} & {\alpha _1^2} &  \cdots  & {\alpha _1^{N - T}}  \\
    \vdots  &  \vdots  &  \vdots  &  \vdots   \\
   {\alpha _{T-1}^1} & {\alpha_{T-1}^2} &  \cdots  & {\alpha_{T-1}^{N - T}}  
\end{array}} \right],\]
i.e., the first downward row rotation. The other parts of the source
are encoded similarly.

The rate of this code can be made arbitrarily close to $1/(N-T)$.
At the decoder, we form a syndrome graph in which there is an
edge between packets $i$ and $j$ (allowing for $j = i$) if there
is an edge between $i$ and $j$ in the syndrome graphs of all of
the layers. For this syndrome graph, delete all nodes without
self-loops, along with their edges. The resulting graph must
have at least one clique of size at least $N-T$, due to the
presence of at least $N-T$ unaltered packets. Thus Lemma~\ref{lemma:claim2} implies 
that there are
at least $N - F(T)$ nodes that are connected to all nodes contained
in a clique of size at least $N - T$. In particular, these $N - F(T)$
nodes must be connected to an unaltered set of nodes of size 
$N - T$. By Lemma~\ref{lem:noloop}, the codewords in all of these $N - F(T)$ 
packets were received correctly.
For each packet, $N-T$ of its layers correspond to
systematic rows of the matrix and $T$ layers correspond to
parities. Thus the decoder can reconstruct a fraction
\[\frac{(N-T)(N-F(T))}{N(N-T)}=\frac {N-F(T)}{N}\]
of the source symbols.


\section{An Impossibility Result}
\label{sec:optimality}
By definition, a polytope code 
\[(f_1,\ldots,f_N,g)\]
is characterized by $(N,T,A,N_0,K_0)$, where $N$ is the number of packets, $T$ is the maximum number of packets that can be altered, $A$ is an eligible $(N,N-T)$-generator matrix, and $N_0$ and $K_0$ are encoding parameters (see Section \ref{sec:Polytope Codes}). From Theorem \ref{the: Main}, we know that for
\[N \ge F(T)+1 \ \ \text{and} \ \ \frac{1}{N-T} \le R \le \frac{1}{N -2T}\]
the R-D pair 
\[\left(R,\frac{F(T)(N-T)(1-(N-2T)R)}{NT}\right)\] 
is achievable using polytope codes. However, when $N \le F(T)$, the decoder in Section~\ref{decoding} no longer works. 

This raises the question of whether our design can be improved when
$N \le F(T)$,
especially since $F(T)$ grows superlinearly with $T$. We next
show the following impossibility result. When $N = F(T)$,
for all sufficiently large $N_0$ and $K_0$, our existing
polytope code construction lacks the partial decodability
property: there exists a set of received packets for which
there is no single packet that can be determined to be 
correct with certainty. Thus, at least as far as 
partial decodability is concerned, neither the decoder
nor the analysis can be improved to relax the 
$N \ge F(T) + 1$ condition; the code itself would need
to change.
Recall that, for polytope codes, in order to drive the rate to 
$1/(N-T)$, we send both $N_0$ and $K_0$ to infinity; see 
(\ref{eq:rateformula}). 
%

To state and prove this result, we use the concept of
\emph {possible transmitted codewords}.
\begin{defn}
\label{def: P_O_C}
Fix $N_0$, $K_0$ and $K$. Given a set of received codewords  $\{\bar y_1^{N_0},...,\bar y_N^{N_0}\}$ and recovered $\{F_{j_1j_2}\}$ for 
$j_1,j_2 \in [N]$ (see Lemma \ref{lem:inner}), if a set of codewords $\{\bar x_1^{N_0},...,\bar x_N^{N_0}\}$ satisfies:
\begin{enumerate}
\item $F_{j_1j_2}= \langle \bar{x}_{j_1}^{N_0},\bar{x}_{j_2}^{N_0} \rangle$, for all $j_1,j_2 \in [N]$;
\item The identity $\bar x_{j}^{N_0} = \bar{y}_{j}^{N_0}$ holds for at 
  least $N - T$ values of $j$ out of $j \in [N]$;
\item $\bar x_{N-T+i}^{N_0} = \sum\nolimits_{j = 1}^{N - T} a_{i,j} {\bar x}_{j}^{N_0}$ for all $i\in [T]$.
\end{enumerate}
then this set of codewords is called a \emph{Possible Transmitted Codeword (PTC)} for $\{\bar y_1^{N_0},...,\bar y_N^{N_0}\}$ and $\{F_{j_1j_2}\}$. Further, let
\begin{multline*}
\mathrm{PTC}(\bar y_{1}^{N_0},...,\bar y_{N}^{N_0},\{F_{j_1j_2}\})=\\
\{\{\bar x_{1,1}^{N_0},...,\bar x_{1,N}^{N_0}\},...,\{\bar x_{M,1}^{N_0},...,\bar x_{M,N}^{N_0}\}\}
\end{multline*}
denote the set of all possible transmitted codewords for $\{\bar y_{1}^{N_0},...,\bar y_{N}^{N_0}\}$ and $\{F_{j_1j_2}\}$. 
\end{defn}

\begin{defn}
\label{def: undecodable}
Fix $N_0$, $K_0$ and $K$ and then fix a set of received packets $\{\bar y_1^{N_0},...,\bar y_N^{N_0}\}$ and recovered $\{F_{j_1j_2}\}$ for $j_1,j_2 \in [N]$. We call $\{\bar y_1^{N_0},...,\bar y_N^{N_0},\{F_{j_1j_2}\}\}$ \emph{totally undecodable} if $\mathrm{PTC}(\bar y_{1}^{N_0},...,\bar y_{N}^{N_0},\{F_{j_1j_2}\})$ has the following property: for any $i\in [N]$, there exists $\{\bar x_{i_1,1}^{N_0},...,\bar x_{i_1,N}^{N_0}\}$ and $\{\bar x_{i_2,1}^{N_0},...,\bar x_{i_2,N}^{N_0}\}$ in $\mathrm{PTC}(\bar y_1^{N_0},...,\bar y_N^{N_0},\{F_{j_1j_2}\})$ such that $\bar x_{i_1,i}^{N_0} \ne \bar x_{i_2,i}^{N_0}$.
\end{defn}

\begin{theorem}
\label{Partial Optimality}
Fix $T>1$, $N=F(T)$ and let $A$ be an $(N,N-T)$ $V$-matrix. Then for all sufficiently large $N_0$ and $K_0$ there exists a set of received packets $\{\bar y_1^{N_0},...,\bar y_N^{N_0}\}$ along with $\{F_{j_1j_2}\}$ such that $\{\bar y_1^{N_0},...,\bar y_N^{N_0},\{F_{j_1j_2}\}\}$ is \emph{totally undecodable}.
\end{theorem}

\begin{IEEEproof}
We begin by showing the conclusion for some $N_0$ and for 
all sufficiently large $K_0$.

Write the $V$-matrix as:\[ A=\left[ {\begin{array}{*{20}{c}}
   1 & 0 &  \cdots  & 0  \\
   0 & 1 &  \ddots  &  \vdots   \\
    \vdots  &  \ddots  &  \ddots  & 0  \\
   0 &  \cdots  & 0 & 1  \\
   {a _{1,1}} & {a _{1,2}} &  \cdots  & {a _{1,N-T}}  \\
    \vdots  &  \vdots  &  \vdots  &  \vdots   \\
   {a _{T,1}} & {a _{T,2}} &  \cdots  & {a _{T,{N-T}}}  \\
\end{array}} \right].\] 
Observe that $\left\lfloor \frac{T}{2} \right\rfloor \left\lceil \frac{T}{2} \right\rceil =N-T-1$. For $i \in \{0,...,\left\lceil \frac{T}{2} \right\rceil-1\}$, let $\mu_i$ denote a length-$\left\lfloor \frac{T}{2} \right\rfloor$ 
integer vector in the right null-space of the $(\left\lfloor \frac{T}{2} \right\rfloor-1 )$-by $\left\lfloor \frac{T}{2} \right\rfloor $ matrix
\begin{equation}
\label{eq:Amatrixpart}
\left[ {\begin{array}{*{20}{c}}
   a_{1, i\left\lfloor \frac{T}{2} \right\rfloor +1}  &  \cdots  & a_{1,(i+1)\left\lfloor \frac{T}{2} \right\rfloor }  \\
   \vdots & \ddots & \vdots \\
   a_{\left\lfloor \frac{T}{2} \right\rfloor -1, i\left\lfloor \frac{T}{2} \right\rfloor+1} & \cdots & a_{\left\lfloor \frac{T}{2} \right\rfloor -1,(i+1)\left\lfloor \frac{T}{2} \right\rfloor } \\
\end{array}} \right].
\end{equation}
Such a vector exists by Lemma~\ref{lem: zero point} in the Appendix
(if $T = 2$, then set $\mu_0 = 1$).
Since $A$ is a $V$-matrix, all 
$(\lfloor \frac{T}{2} \rfloor - 1)$-by-$(\lfloor \frac{T}{2} \rfloor - 1)$
submatrices of the matrix in (\ref{eq:Amatrixpart}) have rank 
$\left\lfloor \frac{T}{2} \right\rfloor-1$ (see Lemma~\ref{lem: support} in 
Appendix~\ref{sec: support}).
Let
$\mu_{i,j}$ refer to the $j$th entry of the column vector $\mu_i$.
Then $\mu_{i,j}$ is non-zero for all $i$ and $j$ by 
Lemma~\ref{lem: zero point}.
For $ i \in \{0,...,\left\lceil \frac{T}{2} \right\rceil-1\}$, let $\nu_i \in \Bbb N^{\left\lfloor \frac{T}{2} \right\rfloor}$ be chosen so that
the components of $\nu_i + \mu_i$ are all positive, then let
\[c_i = [\begin{array}{cc}
 \nu_i & \nu_i+\mu_i 
\end{array} ]\] be an $\Bbb N^{\left\lfloor \frac{T}{2} \right\rfloor \times 2} $ matrix. 
Let $\mu_{\left\lceil \frac{T}{2} \right\rceil}\in \Bbb{Z}$ be a 
natural number whose value will be chosen later, and let
$\nu_{\left\lceil \frac{T}{2} \right\rceil} = 1$. Let
\[c_{\left\lceil \frac{T}{2} \right\rceil} = [\begin{array}{cc}
 \nu_{\left\lceil \frac{T}{2} \right\rceil} & \nu_{\left\lceil \frac{T}{2} \right\rceil}+\mu_{\left\lceil \frac{T}{2} \right\rceil} 
\end{array} ]\]
be an $\Bbb N^{1\times 2}$ matrix.

From $c_i$ define the matrices
\[c_i^+ = [\begin{array}{cc}
 \nu_i +\frac{\mu_i}{2} & \nu_i+\frac{\mu_i}{2}  
\end{array}],\]
and
\[c_i^-=[\begin{array}{cc}
-\frac{\mu_i}{2} & \frac{\mu_i}{2}  
\end{array}].\]
Now let $H$ denote an $L$-by-$L$ Hadamard matrix for some $L$ satisfying
\[L \ge \left\lceil \frac{T}{2} \right\rceil +1,\]
which exists by Sylvester's construction~\cite{sylvester1867lx}.
Each element of $H$ is $-1$ or $1$, and the rows are orthogonal. We use $H$ to construct an $(N-T)$-by-$2L$ matrix $X$ according 
to~(\ref{equ: matrix}).

\begin{figure*}[!t]
\normalsize
\begin{equation}
\label{equ: matrix}
X=\left[\begin{array}{cccc}
 c_0^+ + c_0^- H_{1,1}& c_0^+ + c_0^- H_{1,2} & \cdots &c_0^+ + c_0^- H_{1,L}\\
 c_1^+ + c_1^- H_{2,1}& c_1^+ + c_1^- H_{2,2}& \cdots & c_1^+ + c_1^- H_{2,L}\\
 \vdots& \vdots & \ddots & \vdots\\
c_{\left\lceil \frac{T}{2} \right\rceil-1}^+ + c_{\left\lceil \frac{T}{2} \right\rceil-1}^- H_{\left\lceil \frac{T}{2} \right\rceil,1} &c_{\left\lceil \frac{T}{2} \right\rceil-1}^+ + c_{\left\lceil \frac{T}{2} \right\rceil-1}^- H_{\left\lceil \frac{T}{2} \right\rceil,2} & \cdots&c_{\left\lceil \frac{T}{2} \right\rceil-1}^+ + c_{\left\lceil \frac{T}{2} \right\rceil-1}^- H_{\left\lceil \frac{T}{2} \right\rceil,L}\\
c_{\left\lceil \frac{T}{2} \right\rceil}^+ + c_{\left\lceil \frac{T}{2} \right\rceil}^- H_{\left\lceil \frac{T}{2} \right\rceil+1,1} &c_{\left\lceil \frac{T}{2} \right\rceil}^+ + c_{\left\lceil \frac{T}{2} \right\rceil}^- H_{\left\lceil \frac{T}{2} \right\rceil+1,2}  & \cdots&c_{\left\lceil \frac{T}{2} \right\rceil}^+ + c_{\left\lceil \frac{T}{2} \right\rceil}^- H_{\left\lceil \frac{T}{2} \right\rceil+1,L} \\
\end{array}\right]
\end{equation}
\hrulefill
\vspace*{4pt}
\end{figure*}

Note that for any $i \in \{0,...,\left\lceil \frac{T}{2} \right\rceil\}$, if $H_{i+1,j}=1$,
\[c_i^+ +c_i^- H_{i+1,j} = [\begin{array}{cc}
\nu_i & \nu_i +\mu_i  
\end{array}],\]
and if $H_{i+1,j} = -1$,
\[c_i^+ +c_i^- H_{i+1,j} = [\begin{array}{cc}
\nu_i +\mu_i & \nu_i  
\end{array}].\]
Evidently, the rows of $X$ can be divided into $\left\lceil \frac{T}{2} \right\rceil+1$ blocks, the first 
$\left\lceil \frac{T}{2} \right\rceil$ blocks consisting of $\left\lfloor \frac{T}{2} \right\rfloor$ rows and the last block consisting of a single row.
For $i \in \{0,...,\left\lceil \frac{T}{2} \right\rceil\}$, we define a 
modified version of $X$, $X_i$,
obtained by replacing the $i$th row block in $X$ with
\[[\begin{array}{ccc}
c_i^+ +c_i^- (-H_{i+1,1}) &\cdots &c_i^+ +c_i^- (-H_{i+1,L})  \\
\end{array}].\]
Note that this has the effect of replacing $[\begin{array}{cc}
\nu_i & \nu_i + \mu_i  
\end{array}]$ with $[\begin{array}{cc}
\nu_i + \mu_i & \nu_i  
\end{array}]$ and vice versa. We view $X$ and the various $X_i$ as different source realizations with blocklength $(N-T) N_0 K_0 $ where $N_0 = 2L$
and $K_0$ is any integer satisfying 
$$
\log_K K_0 \ge \max_{i,j} \mu_{i,j} + \nu_{i,j}.
$$
Since $H$ is Hadamard, the inner product between any two rows of $X$ must equal the inner product between the corresponding rows of $X_i$ for all $i$. Thus, all of these source realizations will result in the same norms and inner products being sent as part of the polytope code. Let $\{F_{j_1 j_2}\}$ denote
these norms and inner products.

Next we construct codewords from these source realizations. Let 
\[\bar X = AX\]
and for $i \in \{0,...,\left\lceil \frac{T}{2} \right\rceil\}$, let
\[\bar X_i = AX_i.\]
 
Observe that since $\mu_i$ is in the null space of the matrix in 
(\ref{eq:Amatrixpart}), rows 
\[\left\{N-T+1,...,N-T+ \left\lfloor \frac{T}{2} \right\rfloor-1\}\right\}\] 
of $\bar{X}$ and $\bar{X}_i$ will be the same for all $i \in \{0,..., \left\lceil \frac{T}{2} \right\rceil-1\}$. 

Finally, construct a set of received packets as follows. Packets $1$ through $N-T$ are the first $N-T$ rows of $\bar{X}$, respectively. Packets
\[\left\{N-T+1,...,N-T+ \left\lfloor \frac{T}{2} \right\rfloor-1\right\}\]
are set to be rows $\{N-T+1,...,N-T+ \left\lfloor \frac{T}{2} \right\rfloor-1\}$ of any of the $\bar{X}_i$, $i \in \{0,...,\left\lceil \frac{T}{2} \right\rceil-1\}$ (recall that these rows coincide across $\bar{X}$ and these 
$\bar{X}_i$). For
\[i \in \left\{N-T+\left\lfloor \frac{T}{2} \right\rfloor,...,N\right\},\]
received packet $i$ is set to the corresponding 
row of $\bar{X}_{i-(N-T+\left\lfloor \frac{T}{2} \right\rfloor)}$.
Define the matrix $\bar Y$ to be the set of received packets, one per row, starting with the first.

Now the number of packets that differ between $\bar Y$ and $\bar X_i$ 
is at most 
\[\left\lfloor \frac{T}{2} \right\rfloor + \left\lceil \frac{T}{2} \right\rceil  = T\]
if $i \in \{0,...,\left\lceil \frac{T}{2} \right\rceil-1\}$. Likewise, codeword $\bar X_{\left\lceil \frac{T}{2} \right\rceil}$ differs from $\bar Y$ in at most 
\[1+ \left(\left\lfloor \frac{T}{2} \right\rfloor-1\right)+ \left\lceil \frac{T}{2} \right\rceil = T.\]

Thus,  $\bar{X}_i$, $i \in \{0,...,\left\lceil \frac{T}{2} \right\rceil\}$ 
is in $\mathrm{PTC}(\bar{Y}, \{F_{j_1 j_2}\})$.
For each $i \in \{1,...,N-T\}$, there exists $i_1$ and $i_2$ s.t. row $i$ in $\bar X_{i_1}$ and $\bar X_{i_2}$ disagree. Moreover, we can pick $\mu_{\left\lceil \frac{T}{2}\right\rceil}$ such that for each $i \in \{N-T+1,...,N\}$, row $i$ in $\bar X_1$ and $\bar X_{\left\lceil \frac{T}{2}\right\rceil}$ disagree. This is because for each $i \in \{N-T+1,...,N\}$, there is at most one value for $\mu_{\left\lceil \frac{T}{2}\right\rceil}$ such that row $i$ in $\bar X_1$ and $\bar X_{\left\lceil \frac{T}{2}\right\rceil}$ are the same. Thus the set of integers for which $\mu_{\left\lceil \frac{T}{2}\right\rceil}$ does not satisfy the desired condition has at most $T$ elements, and we 
can choose $\mu_{\left\lceil \frac{T}{2}\right\rceil}$ to be any 
positive integer not in this set.

This establishes the conclusion for $N_0 = 2L$ and 
all sufficiently large $K_0$. One can accommodate larger values
of $N_0$ by prepending a vector of ones to each of the $X_i$
source realizations.
\end{IEEEproof}




\section{Distributed Storage Problem Formulation and Results}\label{sec:dss}
\subsection{Distribution Storage System}

A distributed storage system (DSS) is a collection of storage nodes, each holding a portion of a single data file. We assume each node has capacity $\alpha$, meaning it can store an element of $\mathcal{X}^{n\alpha}$ for some blocklength $n$, where as before $\mathcal{X}=[K]$ is the alphabet set. At any given time, there are $N$ active storage nodes, but individual nodes are unreliable and may fail. When one node fails, a new node is created to replace it. The new node contacts $d$ existing nodes and downloads messages from each one, from which it constructs new storage data. The communication links used to transmit these messages each have capacity $\beta\le\alpha$, meaning they carry elements of $\mathcal{X}^{n\beta}$. The key property that must be maintained is that at any time in this evolution, a data collector (DC) may contact any $k\le d$ existing nodes, download their contents, and perfectly reconstruct the original file. The specific evolution of the system, such as which nodes fail, which nodes are contacted when a new node is formed, and when the DC downloads data to reconstruct the file, is arbitrary and unknown \emph{a priori}. We further assume that there is a finite upper limit $L$ of storage nodes over the lifetime of the storage system (i.e. $N$ initial nodes and at most $L-N$ node failures and replacements), where $L$ is known in advance of code design.\footnote{This is a simplifying assumption not always made in the distributed storage literature, but it is necessary for our results to hold.}  Note that we are considering \emph{functional repair} rather than \emph{exact repair} or \emph{exact repair of systematic parts} (see \cite{DimakisEtal:11IEEE}).

\subsection{Adversary Model}

We assume the presence of an adversary that may take control of a subset of the storage nodes, and alter any message sent from any of those nodes. This includes messages sent when constructing a new node, as well as data downloaded to a DC.  Once a code is fixed, all honest (non-adversarial) nodes behave according to this code, but adversarial nodes may deviate from the code by replacing outgoing transmissions with arbitrary messages. The adversary is omniscient in the sense that it knows the complete stored file, as well as every aspect of the code used by the honest nodes. The adversary may control up to $T$ nodes at any given time. That is, as nodes fail and are replaced, the adversary might continue taking control of new nodes, but at no moment does it control more than $T$ nodes. This is a slightly more pessimistic assumption than in \cite{PawarEtal:11IT}, in which the adversary could control a total of $T$ nodes over the entire evolution of the system, whether or not they existed simultaneously.

We say a rate $R$ is \emph{achievable} for a DSS problem with parameters $(\alpha,\beta,N,k,d,T)$ if for some $n$ there exists a code such that a file $f\in\mathcal{X}^{nR}$ can always be reconstructed without error, no matter the evolution of the system or the adversary actions. The \emph{storage capacity} $C$ is the supremum of all achievable rates. 

\subsection{Bounds on Storage Capacity}

Using a combination of a cut-set bound and the Singleton bound, it was shown in \cite[Theorem 6]{PawarEtal:11IT} that the storage capacity is upper bounded by
\begin{equation}\label{eq:cutset}
	C\le \sum_{i=0}^{k-2T-1} \min\{(d-2T-i)\beta,\alpha\}.
\end{equation}
When $T=0$, the above bound reduces to the exact storage capacity for functional repair without an adversary originally found in \cite{DimakisEtal:10IT}. In other words, this upper bound states that $T$ adversarial nodes yield a storage capacity at most that of the non-adversarial problem with both $d$ and $k$ reduced by $2T$.

Two special points on the storage-bandwidth tradeoff are the so-called Minimum Storage Regenerating (MSR) and Minimum Bandwidth Regenerating (MBR) points. The MSR point is given by
\[
\alpha=(d-k+1)\beta,\quad C=(k-2T)\alpha
\]
and the MBR point is given by
\[
\alpha=(d-2T)\beta,\quad C=\left[(k-2T)(d-2T)-\binom{k-2T}{2}\right]\beta.
\]
In \cite{RashmiEtal:12ISIT}, achievability with exact repair was proved for the MSR point as long as $d-2T\ge 2(k-2T)-2$ and for the MBR point for all parameters, using linear matrix-product codes.

The following theorem is our main achievability result for the distributed storage problem. The proof appears in Section~\ref{sec:dss_proof}.
\begin{theorem}\label{thm:mainbd}
The storage capacity $C$ is lower bounded by
\begin{multline}\label{eq:mainbd}
	C\ge
	 \min\Bigg\{\sum_{i=0}^{k-F(T)-1}\min\{(d-F(T)-i)\beta,\alpha\}, \\
	(d-T)\beta\Bigg\}.
\end{multline}
where $F(T)$ is as  defined in Theorem~\ref{the: Main}.
\end{theorem}

The polytope code used to prove this result, described in detail in Sec.~\ref{sec:dss_proof}, uses a similar decoding procedure to that used for VPEC in Sec.~\ref{decoding} that identifies a subset $V^*$ of trustworthy incoming packets. When constructing a new storage node, this procedure identifies at least $d-F(T)$ trustworthy incoming packets, and when decoding the file at a DC, this procedure identifies at least $k-F(T)$ trustworthy nodes. This explains the first term in \eqref{eq:mainbd}, which corresponds to the capacity of a DSS with no adversary but with $d$ and $k$ each reduced by $F(T)$. The second term in \eqref{eq:mainbd}, limiting the rate to $(d-T)\beta$, ensures that the file could in principle be decoded from the $d-T$ packets sent to a new storage node from honest nodes; this condition ensures that all adversarial packets are either uncorrupted or detected.

\begin{figure}
\includegraphics[width=\columnwidth]{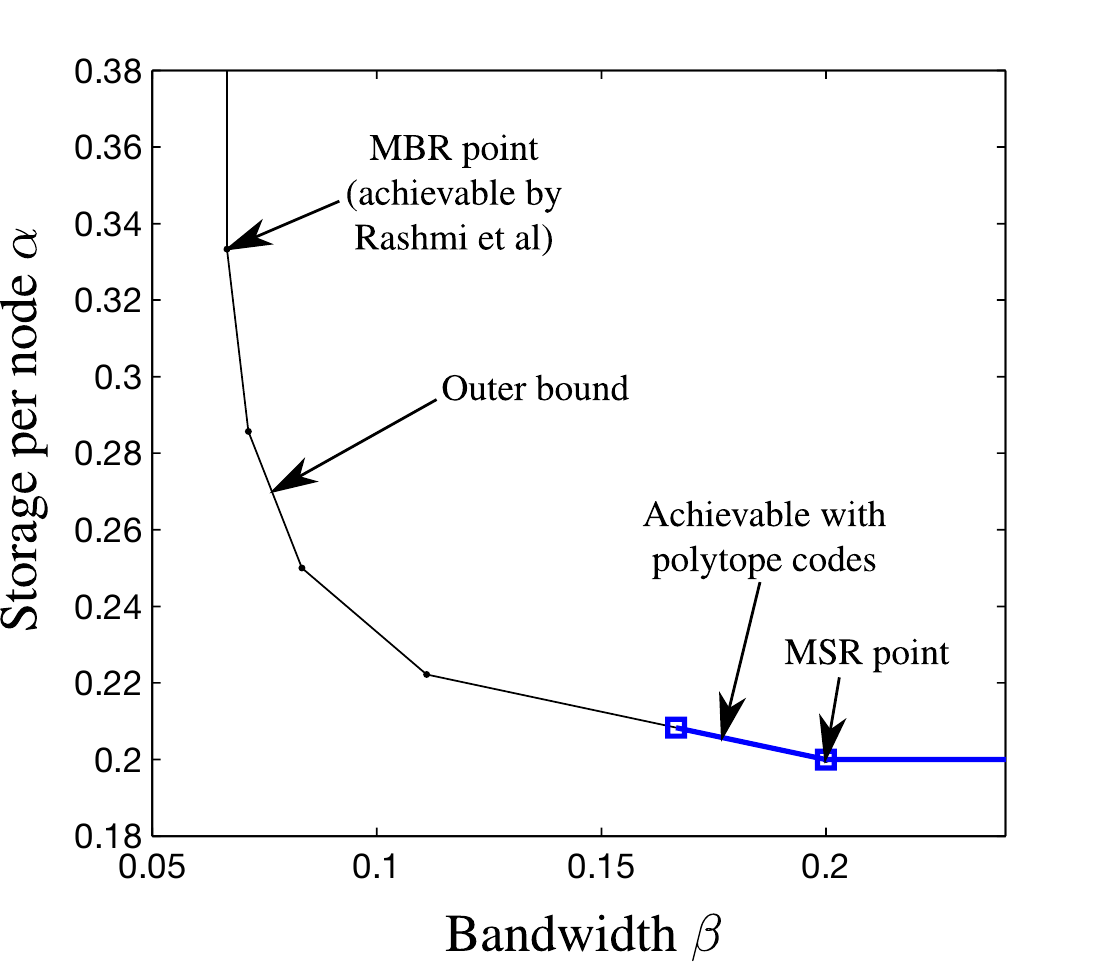}
\caption{Bandwidth-storage tradeoff (i.e. achievable $(\alpha,\beta)$ pairs for $C=1$) for parameters $k=d=7$, $T=1$. Shown is the outer bound \eqref{eq:cutset} found in \cite{PawarEtal:11IT}, and the points achievable with polytope codes by Theorem~\ref{thm:mainbd}. The matrix-product codes from \cite{RashmiEtal:12ISIT} achieve the MBR point, but not the MSR point for these parameters.}
\label{fig:DSSplot}
\end{figure}

Fig.~\ref{fig:DSSplot} illustrates the above bounds on the bandwidth-storage tradeoff (i.e. achievable $(\alpha,\beta)$ for $C=1$) for an example set of parameters. In general, our achievable result matches the upper bound in \eqref{eq:cutset} if $F(T)=2T$ (which holds for $T\le 3$) and the right-hand side of \eqref{eq:cutset} does not exceed $(d-T)\beta$. This includes the MSR point if $T\le 3$ and $(d-2T)(d-k+1)\le d-T$; the latter holds, for example, when $d=k$.

\section{Proof of Theorem~\ref{thm:mainbd}}\label{sec:dss_proof}
We now describe construction of a polytope code to achieve the bound in Theorem~\ref{thm:mainbd}. We assume without loss of generality that $\alpha$ and $\beta$ are integers; if they are not then they can be scaled up and the blocklength $n$ can be scaled down without changing the problem. Let $r$ be the right-hand side of \eqref{eq:mainbd}. We show that rate $r$ can be achieved asymptotically. We fix integers $N_0$ and $K_0$, which play the same roles in the polytope code structure as for the VPEC codes described above. The asymptotic rate $r$ is achieved when both $N_0$ and $K_0$ go to infinity. The file $f$ will be composed of $N_0K_0 r$ symbols from $\mathcal{X}$. The precise blocklength $n$ and rate $R$ will be determined later. We may reparameterize the file as an integer-valued matrix taking values in 
 $\{1,\ldots,K^{K_0}\}^{r\times N_0}$. In particular, we write
\begin{equation}
f=
\left[
\begin{array}{c}
x_{1,K_0}^{N_0} \\
\vdots \\
x_{r,K_0}^{N_0}
\end{array}
\right]
\end{equation}
where $x_{i,K_0}^{N_0}$ is an $N_0$-length vector taking values in $\{1,\ldots, K^{K_0}\}$. As before, we form norms/inner products
\[F_{ij}= \langle x_{i,K_0}^{N_0},x_{j,K_0}^{N_0}\rangle,\forall 1\le i \le j\le r\]
to be included in all packets. We also define for convenience $\mathbf{F}$ to be the vector of all $r+\binom{r}{2}$ norms and inner products.

All packets, both for storage on nodes and for transmissions between nodes, will take the form
\[
(y^{\gamma\times N_0},\mathbf{F},A_0)
\]
where $y^{\gamma\times N_0}$ is a $\gamma\times N_0$ integer-valued matrix, and $A_0$ is a $\gamma\times r$ integer-valued matrix indicating that, with no adversarial influence, we would have
\begin{equation}\label{eq:correct_packet}
y^{\gamma\times N_0}=A_0 f.
\end{equation}
The parameter $\gamma$ represents the size of the data packet: for a storage packet, $\gamma=\alpha$, and for a transmission packet, $\gamma=\beta$.

\emph{Coefficient matrices:} Fix an integer parameter $q$, to be determined later; $q$ plays a role akin to the field size in a code over a finite field, in that it governs the size of the coefficient choices. Let $A$ be a matrix in $\{1,\ldots,q\}^{\alpha N\times r}$ such that any $r\times r$ submatrix of $A$ is nonsingular. The existence of such a matrix for sufficiently large $q$ is guaranteed by Lemma~\ref{lemma:generator}. Now we randomly choose the following coefficient matrices, each independent from the others. For all $1\le i<j\le L$, let $B_{i\to j}$ be a matrix chosen randomly and uniformly from $\{1,\ldots,q\}^{\beta\times\alpha}$. For each $j\in\{n+1,\ldots,L\}$ and each set $V\subseteq\{1,\ldots,j-1\}$ of size at least $d-F(T)$, let $C_{V\to j}$ be a matrix chosen randomly and uniformly from $\{1,\ldots,q\}^{\alpha\times |V|\beta}$. We will prove that for sufficiently large $q$, with positive probability these coefficient matrices yield a code with the required properties, and hence there is at least one successful code.

We now describe operation of the code.

\emph{Data stored on initial nodes:} The initial data to be stored on the $N$ storage nodes is given by
\[
\left[
\begin{array}{c}
y_{1,K_0}^{\alpha\times N_0} \\
\vdots \\
y_{N,K_0}^{\alpha\times N_0}
\end{array}
\right]
  = A f
\]
where $y_{i,K_0}^{\alpha\times N_0}$ is an integer-valued matrix of size $\alpha\times N_0$. On the $i$th storage node, we store packet
\begin{equation}\label{eq:storage_packet}
(y_{i,K_0}^{\alpha\times N_0},\textbf{F},A_i)
\end{equation}
where $A_i$ is the $\alpha\times r$ submatrix of $A$ corresponding to node $i$.

\emph{Transmissions to form new node}: Assume the packet stored on node $i$ is written as in \eqref{eq:storage_packet}. When node $j>i$ is formed, if it contacts node $i$, the packed transmitted from node $i$ to node $j$ is given by
\begin{equation}\label{eq:transmission_packet}
(B_{i\to j} y_{i,K_0}^{\alpha\times N_0},\mathbf{F},B_{i\to j} A_i).
\end{equation}

\emph{Formation of new node}: When node $j$ is formed, the packet it stores is formed as follows. Node $j$ first determines $\mathbf{F}$ using majority rule among all its received packets. Then it uses the procedure described in Sec.~\ref{decoding} to find a set $V^*_j\subset\{1,\ldots,j-1\}$ of trustworthy incoming packets. By Lemma~\ref{lemma:claim2}, $|V^*_j|\ge d-F(T)$. Let $z_{j,K_0}^{|V^*_j|\beta\times N_0}$ be the $|V^*|\beta\times N_0$ matrix composed of the data stored in these trustworthy packets, and let $A_{\to j}$ be the concatenation of the corresponding coefficient matrices. The packet stored at node $j$ is then given by
\[
(C_{V^*_j\to j} z_{j,K_0}^{|V^*_j|\beta\times N_0}, \mathbf{F},C_{V^*_j\to j} A_{\to j}).
\]

\emph{Decoding at a data collector}: To decode the original message, the DC downloads the packets stored on $k$ nodes. After recovering $\mathbf{F}$ using majority rule, it again uses the procedure in Sec.~\ref{decoding} to find a set $V^*_{\text{DC}}$ of trustworthy incoming packets, where $|V^*_{\text{DC}}|\ge k-F(T)$. Let $z^{|V^*_{\text{DC}}|\alpha\times N_0}_{K_0}$ be the concatenation of the data matrices on these packets, and $\hat{A}$ be the concatenation of the corresponding coefficient matrices. The DC declares its estimate $\hat{f}$ to be the unique $r\times N_0$ matrix such that
\begin{equation}\label{eq:file_decode}
z^{|V^*_{\text{DC}}|\alpha\times N_0} = \hat{A} \hat{f}.
\end{equation}
If there is no such value or more than one, declare an error.

\emph{Rate analysis}: First note that $|F_{ij}|\le K^{2K_0}N_0$, so the number of symbols required to store $\mathbf{F}$ is at most 
\[
(2K_0+\log_K K_0)\left[r+\binom{r}{2}\right].
\]
Next we bound the coefficient matrices $A_i$. By construction, for $i=1,\ldots,N$, the each element of $A_i$ is in $\{1,\ldots,q\}$. We prove by induction that, for all $j=N+1,\ldots,L$, each element of $A_j$ is a positive integer no more than
\[
(q^2 \alpha\beta d)^{j-N} q.
\]
Indeed, assume that for all $i<j$, each element of $A_i$ is at most
\[
(q^2\alpha\beta d)^{i-L}q\le (q^2 \alpha\beta d)^{j-N-1} q.
\]
Thus, each element of matrix $B_{i\to j}A_i$ (and hence each element of $A_{\to j}$) is at most
\[
(q\alpha) (q^2\alpha\beta d)^{j-N-1}q.
\]
Since $A_j=C_{V_j^*\to j} A_{\to j}$, and $C_{V_j^*\to j}\in \{1,\ldots, q\}^{\alpha\times|V^*_j|\beta}$ where $|V_j^*|\le d$, each element of $A_j$ is at most
\[
(q \beta d)(q\alpha)(q^2\alpha\beta d)^{j-N-1}q
= (q^2\alpha\beta d)^{j-N}q.
\]
Therefore, for all nodes $i=1,\ldots,L$, the elements of $A_i$ are at most
\[
(q^2\alpha\beta d)^{L-N}q.
\]
Thus the elements of $y_{i,K_0}^{\alpha\times N_0}$ are at most
\[
(q^2\alpha\beta d)^{L-N}q r K^{K_0}.
\]
Thus to store $y_{i,K_0}^{\alpha\times N_0}$ requires
\[
\alpha N_0(K_0+\lceil\log_K (q^2\alpha\beta d)^{L-N} qr\rceil)
\]
symbols, and to store $A_i$ requires
\[
\alpha r \lceil\log_K (q^2\alpha\beta d)^{L-N}q\rceil
\]
symbols. The total number of symbols stored on node each node in the packet \eqref{eq:storage_packet} is therefore
\begin{multline*}
(2K_0+\log_K K_0)\left[r+\binom{r}{2}\right]+\alpha r \lceil\log_K (q^2\alpha\beta d)^{L-N}q\rceil\\+\alpha N_0(K_0+\lceil\log_K (q^2\alpha\beta d)^{L-N} qr\rceil).
\end{multline*}
Similarly, the total number of symbols transmitted from one node to another in the packet \eqref{eq:transmission_packet} is at most
\begin{multline*}
(2K_0+\log_K K_0)\left[r+\binom{r}{2}\right]+\beta r \lceil\log_K (q^2\alpha\beta d)^{L-N}q\rceil\\+\beta N_0(K_0+\lceil\log_K (q^2\alpha\beta d)^{L-N} qr\rceil).
\end{multline*}
Since $\beta\le\alpha$, taking the blocklength to be
\begin{multline*}
n=\frac{1}{\beta} (2K_0+\log_K K_0)\left[r+\binom{r}{2}\right]+ r \lceil\log_K (q^2\alpha\beta d)^{L-N}q\rceil\\+ N_0(K_0+\lceil\log_K (q^2\alpha\beta d)^{L-N} qr\rceil)
\end{multline*}
allows us to form the storage packets as  $n\alpha$ symbols and the transmission packets as $n\beta$ symbols. Since the file is given by $N_0K_0 r$ symbols, the rate achieved by this code is
\[
R=\frac{N_0K_0 r}{n}
\]
which may be made arbitrarily close to $r$ for sufficiently large $N_0$ and $K_0$.

\emph{Proof of correctness}: The following lemma is proved below.

\begin{lemma}\label{lem:rank}
For sufficiently large $q$, which positive probability on the choice of coefficient matrices $B_{i\to j}$ and $C_{V\to j}$, the following hold:
\begin{enumerate}
\item for any DC, the corresponding coefficient matrix $\hat{A}$ has rank $r$,
\item for each node $j$, the matrix $\bar{A}_{\to j}$, consisting of the rows of $A_{\to j}$ corresponding to the honest nodes, has rank $r$.
\end{enumerate}
\end{lemma}

We first prove that no honest storage nodes ever stores faulty data. That is, \eqref{eq:correct_packet} always holds for stored packets at honest nodes. By construction, the initial honest nodes store only truthful data. We proceed by induction: assume all existing honest nodes hold truthful data, and we show that when a new node $j$ is formed, all packets sent from nodes in $V^*$ hold truthful data, even if sent by an adversarial node. There must be at least $d-T$ honest nodes that transmit packets, which, by the inductive hypothesis, all send truthful packets. Thus these $d-T$ nodes form a clique in the syndrome graph. Thus, for any adversarial node $i\in V^*_j$, the syndrome graph must include a self-loop, as well as an edge from $i$ to each of these $d-T$ honest nodes. Moreover, by Lemma~\ref{lem:rank}, matrix $\bar{A}_{\to j}$ has rank $r$; in other words, the entire message can be determined from the packets sent from honest nodes. Thus the unaltered data for any node $i\in V^*_j$ is a linear combination of the data sent from honest nodes. Therefore, by Lemma~\ref{lem:noloop}, the packet from $i$ to $j$ is unaltered.

Now we show that the DC always decodes correctly. As we have proved, all honest nodes store only truthful data. Thus, when the DC downloads data from $k$ nodes, at least $k-T$ of them contain only truthful data. By a similar argument as above, any node in $V^*_{\text{DC}}$ contains truthful data. Since by Lemma~\ref{lem:rank} matrix $\hat{A}$ has rank $r$, the only value $\hat{f}$ satisfying \eqref{eq:file_decode} is the true value of the file $f$.

\begin{IEEEproof}[Proof of Lemma~\ref{lem:rank}]
We make use of the \emph{information flow graph} developed in \cite{DimakisEtal:10IT}. The basic insight is that the distributed storage problem can be posed as a multicast network coding problem on the information flow graph, described as follows. The graph, denoted $G_{\text{DSS}}$, consists of a source node $\mathsf{S}$, for each storage node $i$ a pair of nodes $\mathsf{x}_\text{in}^i$ and $\mathsf{x}_\text{out}^i$, and for each DC a node $\mathsf{DC}_j$. Each pair of storage nodes are connected by a link $\mathsf{x}_\text{in}^i\to\mathsf{x}_\text{out}^i$ of capacity $\alpha$. For the initial storage nodes $j=1,\ldots,N$, there is a link $\mathsf{S}\to \mathsf{x}_\text{in}^i$ of infinite capacity. For subsequent storage nodes $j>N$, there is a link $\mathsf{x}_\text{out}^i\to \mathsf{x}_\text{in}^j$ of capacity $\beta$ for each of the $d$ nodes $i$ that transmit a message to node $j$. For each data collector, there is a link $\mathsf{x}_\text{out}^i\to \mathsf{DC}_j$ of infinite capacity for each of the $k$ nodes $i$ from which the DC downloads data. It is shown in \cite[Lemma~2]{DimakisEtal:10IT} that for any DC, the min-cut of this graph from the source $\mathsf{S}$ to $\mathsf{DC}_j$ is lower bounded by
\[
\sum_{i=0}^{k-1} \min\{(d-i)\beta,\alpha\}.
\]

Consider the subgraph $\tilde{G}_{\text{DSS}}$ of the information flow graph in which, for each node $j>N$, the links incoming to $\mathsf{x}_\text{in}^j$ from nodes not in $V^*_j$ are deleted, and similarly links to the DC not in $V^*_{\text{DC}}$ are deleted. Note that, on this subgraph, the polytope code behaves essentially like an ordinary linear network code without adversaries, except that linear operations are over the integers rather than a finite field. We further define, for each node $i>N$, a different subgraph $\tilde{G}^{(i)}_{\text{DSS}}$ of the information flow graph, which is the same as $\tilde{G}_{\text{DSS}}$ except that all incoming links to $\textsf{x}_\text{in}^i$ from honest nodes are retained.

By standard arguments in linear network coding (see, for example, \cite{HoEtal:06IT}), which apply equally well for integer operations as for a finite field, for sufficiently large $q$, with probability approaching $1$, the rank of a coefficient matrix will be equal to the min-cut of the corresponding information flow graph. Therefore, to prove the lemma it is enough to prove the following two min-cut properties:
\begin{enumerate}
\item On $\tilde{G}_{\text{DSS}}$, the min-cut from $\textsf{S}$ to $\textsf{DC}_j$ for any $j$ is at least $r$.
\item On $\tilde{G}^{(i)}_{\text{DSS}}$, the min-cut from $\textsf{S}$ to $\textsf{x}_{\text{in}}^j$ is at least $r$.
\end{enumerate}

The first of these properties is easily proved using existing information flow results. In particular, since $|V^*_j|\ge d-F(T)$ and $|V^*_{\text{DC}}|\ge k-F(T)$, we may apply \cite[Lemma~2]{DimakisEtal:10IT} to find that the min-cut on $\tilde{G}_{\text{DSS}}$ from $\mathsf{S}$ to $\mathsf{DC}_j$ is lower bounded by
\[
\sum_{i=0}^{k-F(T)-1} \min\{(d-F(T)-i)\beta,\alpha\}\ge r.
\]

The proof of the second min-cut property requires a slight modification of that of \cite[Lemma 2]{DimakisEtal:10IT}. Let $(U,\bar{U})$ be any cut on $\tilde{G}_\text{DSS}^{(i)}$ where $\mathsf{S}\in U$ and $\textsf{x}_\text{in}^i\in \bar{U}$. Let $\mathcal{C}$ be the set of edges connecting $U$ to $\bar{U}$. Let $z$ be the number of output nodes in $\bar{U}$. Let $\textsf{x}_\text{out}^{j_1}$ be the first such node in $\bar{U}$. There are two cases:
\begin{itemize}
\item If $\textsf{x}_\text{in}^{j_1}\in U$, then the edge $\textsf{x}_\text{in}^{j_1}\to \textsf{x}_\text{in}^{j_1}$ is in $\mathcal{C}$.
\item If $\textsf{x}_\text{in}^{j_1}\in \bar{U}$, then the incoming edges to $\textsf{x}_\text{in}^{j_1}$, all of which come from output nodes in $U$, are in $\mathcal{C}$. There are at least $d-F(T)$ of these edges.
\end{itemize}
These edges contribute at least $\min\{(d-F(T))\beta,\alpha\}$ to the cut capacity.

Let $\textsf{x}_\text{out}^{j_2}$ be the next output node in $\bar{U}$. Again there are two cases:
\begin{itemize}
\item If $\textsf{x}_\text{in}^{j_2}\in U$, then the edge $\textsf{x}_\text{in}^{j_2}\to \textsf{x}_\text{in}^{j_2}$ is in $\mathcal{C}$.
\item If $\textsf{x}_\text{in}^{j_2}\in\bar{U}$, since only one edge incoming to $\textsf{x}_\text{in}^{j_2}$ may come from $\textsf{x}_\text{out}^{j_1}$, at least $d-F(T)-1$ of its incoming edges are in $\mathcal{C}$.
\end{itemize}
These edges contribute at least $\min\{(d-F(T)-1)\beta,\alpha\}$ to the cut capacity. Continuing this reasoning, we  accumulate a total cut capacity of
\[
\sum_{i=0}^{\min\{z-1,d-F(T)\}} \min\{(d-F(T)-i)\beta,\alpha\}.
\]
In addition, since $\textsf{x}_\text{in}^i$ has at least $d-T$ incoming edges, if $z<d-T$ then at least $d-T-z$ incoming edges to $\textsf{x}_\text{in}^i$ are in $\mathcal{C}$. Thus, the total cut capacity is at least
\begin{equation}\label{eq:mincut}
\sum_{i=0}^{\min\{z,d-F(T)\}-1} \min\{(d-F(T)-i)\beta,\alpha\}+|d-T-z|^+\beta.
\end{equation}
If $z\le d-F(T)$, then since $F(T)\ge T$ we have $z\le d-T$, so \eqref{eq:mincut} is at least
\begin{align*}
&\sum_{i=0}^{z-1} \min\{(d-F(T)-i)\beta,\alpha\}+(d-T-z)\beta
\\&\ge \sum_{i=0}^{z-1} \beta+(d-T-z)\beta 
\\&= (d-T)\beta
\ge r.
\end{align*}
If $z >d-F(T)$, then \eqref{eq:mincut} is at least
\begin{align*}
&\sum_{i=0}^{d-F(T)-1} \min\{(d-F(T)-i)\beta,\alpha\}
\\&\ge \sum_{i=0}^{k-F(T)-1} \min\{(d-F(T)-i)\beta,\alpha\}
\\&\ge r.
\end{align*}
Therefore, in any case the min-cut from $\textsf{S}$ to $\textsf{x}_\text{in}^i$ is at least $r$.

\end{IEEEproof}

\appendices

\section{Supporting Lemmas}
\label{sec: support}
\begin{lemma}
\label{lem: zero point}
For any integer $m >1$ and $\Lambda \in \Bbb Z ^{(m-1)\times m}$, there exists a non-zero vector $x^m \in \Bbb Z ^m$ such that $\Lambda x^m = 0$. Furthermore, if $\mathrm{rank}(\Lambda')=m-1$ for all $(m-1)$-by-$(m-1)$ submatrices
$\Lambda'$ of $\Lambda$, then any such an $x^m$ must be in
$(\Bbb Z \backslash \{0\} ) ^m$.
\end{lemma}
\begin{IEEEproof}
Let $\lambda_1^m,...,\lambda_{m-1}^m$ denote the rows of $\Lambda$. Using the Gram-Schmidt procedure, we may assume that $\lambda_1^m,...,\lambda_{m-1}^m$ are orthogonal. Since $\lambda_1^m,...,\lambda_{m-1}^m$ cannot span $\Bbb R ^m$ but $\Bbb N^m$ does, there must exist a vector $\lambda^m \in \Bbb N^m$ that is not in the span of $\lambda_1,...,\lambda_{m-1}$. Then the vector:
\[\lambda^m - \sum_{i=1}^{m-1} \frac{(\lambda_i^m)^T \lambda^m}{(\lambda_i^m)^T\lambda^m_i}\lambda_i^m,\]
where the sum excludes those $i$ for which $\lambda_i^m$ is the zero vector,
is in $\Bbb Q^m$ and is orthogonal to $\lambda_1^m,...,\lambda^m_{m-1}$. Multiplying $\lambda^m$ by the least common denominator gives a non-zero integer solution to $\Lambda x^m=0$.

When $\mathrm{rank}(\Lambda')=m-1$ for all $\Lambda'$,
we prove that all the entries of $x^m$ must be non-zero by contradiction. Without loss of generality, suppose that $x_1=0$. Then
\[[\begin{array}{ccc}
 \Lambda_2& \cdots & \Lambda_{m}  
\end{array}] \left[\begin{array}{c}
 x_2 \\
 \vdots \\
 x_{m}
\end{array}\right]=0,\]
where $\Lambda_2$ through $\Lambda_{m}$ are the second through
last columns of $\Lambda$. Now
$[\begin{array}{ccc}
 \Lambda_2& \cdots & \Lambda_{m}  
\end{array}]$ is a non-singular matrix by hypothesis. The above
linear system then has a unique solution, namely the zero
vector. This implies that $x^m$ is the zero vector, which is a contradiction.
\end{IEEEproof}

\begin{lemma}
\label{lem: support}
Let $\alpha_1, \ldots \alpha_m$ be distinct natural numbers. Then
for any integer $k \ge 0$, every $m$-by-$m$ submatrix of
$$
M = \left[\begin{array}{cccc}
\alpha_1^k & \alpha_1^{k+1} & \cdots & \alpha_1^{k+m} \\
\alpha_2^k & \alpha_2^{k+1} & \cdots & \alpha_2^{k+m} \\
\vdots & \vdots & \ddots & \vdots \\
\alpha_m^k & \alpha_m^{k+1} & \cdots & \alpha_m^{k+m} \end{array}\right].
$$
is nonsingular.
\end{lemma}

\begin{IEEEproof}
Let $a = [a_0 \ \ a_1 \ \cdots \ a_m]^T$ be such 
that $M a = 0$ and $a_i = 0$ for some $i$. It 
suffices to show that $a$ must be the zero vector. Now $a$ is
in the nullspace of 
$$
M = \left[\begin{array}{cccc}
1 & \alpha_1^{1} & \cdots & \alpha_1^{m} \\
1 & \alpha_2^{1} & \cdots & \alpha_2^{m} \\
\vdots & \vdots & \ddots & \vdots \\
1 & \alpha_m^{1} & \cdots & \alpha_m^{m} \end{array}\right].
$$
Consider the polynomial
$$
P(x) = \sum_{i = 0}^m a_i x^i.
$$
Evidently $P$ is a degree-$m$ polynomial with roots 
$\alpha_1$, \ldots, $\alpha_m$. There is a unique nonzero
degree-$m$ polynomial with these roots, however, namely, 
$$
P'(x) = \prod_{i = 0}^m (x - \alpha_i) = \sum_{i = 0}^m a_i' x^i.
$$
Since all of the $\alpha_i$ are positive, all of the $a_i'$ must
be nonzero. It follows that $P(\cdot) \ne P'(\cdot)$ and so
$P(\cdot)$ must be the all-zero polynomial.
\end{IEEEproof}

\section*{Acknowledgment}

The authors wish to thank Ebad Ahmed for his contributions
to this work during its early stages. The scheme in (\ref{eq:ebad3packet}),
in particular, is due to him.
This research was
supported by the Army Research Office under grant W911NF-13-1-0455
and by the National Science Foundation under 
{grants CCF-1117128, CCF-1218578, and CCF-1453718.}



%
\bibliographystyle{plain}
\bibliography{mybib,KosutBibs}

\end{document}